\documentclass[12pt,a4,english]{article}
\usepackage[T1]{fontenc}
\usepackage[latin9]{inputenc}
\usepackage{geometry}
\usepackage{verbatim}
\usepackage{amsmath}
\usepackage{graphicx}
\usepackage{longtable}
\usepackage{setspace}
\usepackage[round,sort,authoryear]{natbib}
\usepackage{amsfonts}
\usepackage{graphicx}
\usepackage{babel}
\usepackage{amsthm}
\usepackage{booktabs,caption}
\usepackage{enumitem}
\usepackage{amsfonts}
\usepackage{accents}
\usepackage{natbib}
\usepackage{bibentry}
\usepackage{amsmath}
\usepackage{mathtools}
\usepackage[colorlinks=,linkcolor=blue, citecolor=blue, ]{hyperref}
\usepackage{pdflscape}
\usepackage{cmbright}
\usepackage[title]{appendix}


\makeatletter
\newcommand{\nextverbatimspread}[1]{  \def\verbatim@font{    \linespread{#1}\normalfont\ttfamily    \gdef\verbatim@font{\normalfont\ttfamily}}}
\makeatother
\nobibliography*

\DeclarePairedDelimiter{\floor}{\lfloor}{\rfloor}
\usepackage[T1]{fontenc}

\theoremstyle{definition}
\newtheorem{theorem}{\normalfont\scshape \bfseries Theorem}

\newtheorem{assumption}{\normalfont\scshape \bfseries Assumption}
\newtheorem{lemma}{\normalfont\scshape \bfseries Lemma}

\newtheorem{proposition}{\normalfont\scshape \bfseries Proposition}
\newtheorem{remark}{\normalfont\scshape \bfseries Remark}

\newtheorem{corollary}{\normalfont\scshape \bfseries Corollary}

\newcommand\norm[1]{\left\lVert#1\right\rVert}

\numberwithin{equation}{section}

\AtBeginDocument{}

\begin{document}

\pagenumbering{roman}












\title{ {\Large \textbf{Predictability Tests Robust against Parameter Instability}
}
}
\author{Christis Katsouris\thanks{PhD Candidate, Department of Economics, University of Southampton - Highfield Campus, UK. 
\textit{Email:} \textcolor{blue}{C.Katsouris@soton.ac.uk}. I am grateful to my academic advisors Jean-Yves Pitarakis, Jose Olmo and Tassos Magalinos for their guidance, support and continuous encouragement. Moreover, I would like to thank session participants of the IX Workshop in Time Series Econometrics at the University of Zaragoza, the 2019 Conference of the International Association of Applied Econometrics (IAEE) at the University of Cyprus, as well as the participants of the 2018 and 2019 Econometric Workshops at the Department of Economics, for helpful conversations. Financial support from the University of Southampton Vice Chancellors PhD Scholarship is gratefully acknowledged. The author also acknowledge the use of the IRIDIS High Performance Computing Facility, and associated support services at the University of Southampton, in the completion of this work. All the usual disclaimers apply.  } 
\\
\small{Department of Economics, University of Southampton}
\\
\textcolor{blue}{Job Market Paper I} \textcolor{blue}{(unedited)} 
}
\date{January 25, 2021}

\maketitle

\begin{center}
\textbf{Abstract}
\end{center}
\vspace{-0.8\baselineskip}
We consider Wald type statistics designed for joint  predictability and structural break testing based on the instrumentation method of \cite{phillipsmagdal2009econometric}. We show that under the assumption of nonstationary predictors: (i) the tests based on the OLS estimators converge to a nonstandard limiting distribution which depends on the nuisance coefficient of persistence; and (ii) the tests based on the IVX estimators can filter out the persistence under certain parameter restrictions due to the supremum functional. These results contribute to the literature of joint predictability and parameter instability testing by providing analytical tractable asymptotic theory when taking into account nonstationary regressors. We compare the finite-sample size and power performance of the Wald tests under both estimators via extensive Monte Carlo experiments. Critical values are computed using standard bootstrap inference methodologies. We illustrate the  usefulness of the proposed framework to test for predictability under the presence of parameter instability by examining the stock market predictability puzzle for the US equity premium.       
\\

\textit{JEL classification:} C12, C22, C26, C32, C53, C58. 

\textbf{Keywords:}  Stock Return Predictability, Parameter Instability, Predictive Regression, IVX filtration, Local-to-unity asymptotic theory, Ornstein-Uhlenbeck process, bootstrap.    






\newpage 

\setcounter{page}{1}
\pagenumbering{arabic}






\section{Introduction}

The stock predictability puzzle is an important research theme appeared in the financial econometrics literature which has seen growing attention in recent years\footnote{In empirical finance the predictive regression model provides a suitable testing framework for aspects such as conditional asset pricing and investment strategy performance evaluation. An extensive study about the latter is presented by \cite{pesaran1995predictability}. The author examines the predictability of stock returns from the perspective of optimal portfolio decision. Moreover, the author introduces the concept of episodic predictability, arguing that predictability can be explained in accordance to certain macroeconomic events. Moreover, related applied macroeconometric applications include the forecasting of macroeconomic fundamentals as well as the examination of short-horizon vis-a-vis long-horizon predictability.}. Various studies have  demonstrated that forecast performance via predictive regressions with macroeconomic variables as predictors, is countercyclical, that is, there is a co-movement with the business cycle. Further studies focused on identifying periods of episodic predictability; a terminology adapted to describe the phenomenon of predictors "switching on" during certain periods, while appear to have no predictive ability during other periods (e.g., see \cite{gonzalo2012regime, gonzalo2017inferring}, \cite{chinco2019sparse}, \cite{demetrescu2020testing}). These periods of unstable forecasting ability appear in time series models in the form of parameter instability (see, \cite{rossi2012out},  \cite{inoue2017rolling}, \cite{pitarakis2017simple} and \cite{georgiev2018testing}). Considering this vibrant discussion, we aim to study how the degree of persistence affects the asymptotic theory of related to the above aspects. Our goal is to tackle the following econometric question: "How does the time series properties of predictors affect predictability testing under the presence of parameter instability?"

The predictive regression model operates under the strong assumption of parameter stability, which could be violated in certain regions of the  sample\footnote{A related study to parameter instability in prediction models is presented by \cite{paye2006instability}. In this paper, the authors propose a framework for identifying and estimating multiple breaks.}. Moreover, the majority of the structural break literature propose testing methodologies under the assumption of stationary time series. Therefore, the development of a robust joint test against both predictability and structural break is an important aspect to tackle. Related issues from the predictability literature, include the high persistence of predictors as well as the econometric implications of testing for structural break given the existence of nonstationary regressors as captured via the local-to-unit root specification of the predictive regression.
In this paper we propose an econometric framework for jointly testing against both predictability and structural break. The tests are constructed in a similar manner as the Wald type statistics proposed by \cite{gonzalo2012regime} in a threshold predictive regression model\footnote{The particular framework and in the subsequent paper of \cite{gonzalo2017inferring} the authors propose tests which capture the effects of linearity and the presence of a threshold effect in predictive regressions with persistent predictors.}.  Our contributions are threefold: (i) We propose a test statistic which jointly tests against the alternative hypothesis of both predictability and structural break and show that the test is robust to the persistence properties of predictors in the model; 

\newpage 

(ii) We show that the test statistic has a nuisance parameter free limiting distribution under the assumption of stationary and mildly stationary regressors, while it converges to a nonpivotal asymptotic distribution under the assumption of local-to-unit root (LUR) or integrated regressors; (iii) We provide an extensive Monte Carlo simulation study where we compare the empirical size and power of the sup IVX-Wald test. We also present simulations to examine the finite-sample properties of the corresponding sup OLS-Wald statistic. Lastly, we employ the proposed testing framework to investigate the predictability puzzle based on the equity premium of US stock returns.  


\paragraph{Related Literature} 

The ideas in this paper are related to research done in two different fields. From finance, it is related to the stock return predictability literature and from the econometrics and statistics perspective it is related to the literature of parameter instability and structural break testing methodologies. We aim to bridge the gap in the literature by proposing a testing framework which incorporates both aspects.   

Firstly, we provide a general overview of the predictability literature, starting with the implementation of simple t-tests to detect statistical significance in stable relations of predictants such as equity index returns, on the lagged time series of predictors. From extensive empirical applications, this practise has been proved to cause distorted inference due to the presence of high persistent predictors, since nonstandard terms appear in the limiting distribution of the t-test. In particular, \cite{stambaugh1999predictive} observed this finite sample bias\footnote{Notice that the Stambaugh bias correction is based on the studies of \cite{marriott1954bias} and \cite{kendall1954note} who proposed suitable bias correction in autocorrelations.} which occurs when the classical least squares estimator is employed for statistical inference. Furthermore, \cite{amihud2004predictive} consider a second-order bias-correction and propose a reduced-bias OLS based estimator. Both aforementioned methods are considered to provide a post-estimation bias correction in finite samples.  

Secondly, the predictability literature was extended to nonstandard inference to account for the presence of nonstationary predictors; some of the most notable contributions include the Bonferroni-type approach as in  \cite{cavanagh1995inference} and \cite{campbell2006efficient}, the conditional likelihood method based on sufficient statistics\footnote{The idea of sufficient statistics considers optimal tests invariant under transformation based on the curved exponential family. This framework allows to use the conditional restrictions testing problem in the presence of nuisance parameters and is particularly appealing in the case of near integrated regressors.} proposed by \cite{jansson2006optimal} and the control function method proposed by \cite{elliott2011control}. The drawback of these testing methodologies is that the asymptotic   theory has some undesirable properties such as the uncorrectable bias due to the presence of nuisance parameter in the limit distribution (see, \cite{phillips2013predictive}). Moreover, the particular approaches are computational complex especially in multivariate predictive regression settings. For instance, \cite{kasparis2015nonparametric} propose a set of nonparametric predictive tests which allows for nonlinearities and offers robustness to integration order.

\newpage 

Thirdly, a novel approach that recently has been attracting much attention is the instrumental variable-based test proposed by \cite{kostakis2015Robust} (KMS, hereafter), which is build upon the theoretical framework developed by \cite{phillipsmagdal2009econometric}. This methodology, referred to as IVX-Wald test, provides a robust framework for predictive regression models which is valid for predictors with general persistence properties. Specifically, the asymptotic theory shows that the IVX estimator which converges to a mixed Gaussian distribution, successfully removes the long-run endogeneity, that appears due to the innovation structure of the model, and provides a pivotal statistic robust under different degrees of persistence or even regressors of mixed integration (e.g., see \cite{phillips2016robust}). Hence, a self-normalized Wald statistic can be constructed that converges to a nuisance parameter free $\chi^2$ limiting distribution. More importantly, the IVX filtration implies a direct inference procedure via the various moment approximations (e.g., long-run covariance matrices) and can be easily extended to the multivariate predictive regression model under certain regulatory conditions.   

The aforementioned literature operates under the assumption of parameter constancy\footnote{Note that, the literature of structural change goes back to 1940s and 1950s with the pioneering work of Wald on sequential hypothesis testing as well as the seminal work of \cite{page1954continuous} which proposes methodologies for detecting anomalies in control charting, an idea further developed by \cite{chu1995moving} and \cite{chu1996monitoring} who consider testing for structural break in the sense of contaminated and non-contaminated periods in linear regression models. Further seminal studies for structural break tests include \cite{chow1960tests}, \cite{hawkins1987test} and \cite{ploberger1990local} among others.} which implies a stable predictive relationship over the sample period. However, due to the nature of economic conditions, shifting between periods of market  tranquillity and periods of market exuberance, the phenomenon of episodic predictability\footnote{Note that, formal econometric methods to test for episodic predictability have been recently proposed by \cite{demetrescu2020testing}, who consider related subsampling techniques.} has been proposed to capture these "pockets" of predictability across business cycles (see, \cite{farmer2019pockets}). This implies the existence of time-varying predictability which can be examined within an econometric framework which accommodates time-varying parameters. In this paper, we approach this aspect by proposing to test for the existence of joint predictability and structural break, in the form of a single parameter shift at an unknown break-point. Thus, our approach is closer to the framework of \cite{andrews1993tests} who proposed Wald-type statistics for testing for a single structural break at an unknown break-point. 

The presence of structural break in predictive regression models implies variation in predictability across time. As a result, standard testing methodologies for break detection, such as Andrew's Wald statistics for linear regression models as well as the family of tests proposed by \cite{bai1998estimating} can no longer be valid for nonstationary\footnote{The frameworks proposed by \cite{phillipsmagdal2009econometric} and \cite{magdalinos2009limit} explicitly examine the various forms of nonstationarity attributed due to the structure of the predictive and cointegrated system, as captured by specific characteristics of the system. This is a major distinction in the literature which previously thought that the presence of nonstationarity in the regressors as a charateristic occurring due to the autoregressive specification of the model predictors solely.} predictors  

\newpage 

as captured by the autoregressive specification of the predictive regression model. This invalidity is demonstrated by the simulation studies of \cite{paye2006instability}. More specifically, the authors show that in the case of highly correlated innovations and (unfiltered) persistent predictors, an implementation of a sup-F test and UDMax test can cause severe size distortions when testing for structural breaks in time series. Moreover, the optimal test of Elliot and Muller has good empirical size performance, but distortions appear in the asymptotic properties of the power function of the test.

The literature of structural break tests has recently focused in the construction of suitable testing frameworks for predictive regression models. \cite{cai2015testing} propose modelling smooth structural breaks using an $L_2-$ type statistic, in predictive regressions with nonstationary predictors. A different approach include the study of \cite{pitarakis2017simple} who consider a CUSUM-type statistic. Additionally, due to the occurrence of nonstandard limiting distributions  in standard structural break test implementations, \cite{georgiev2018testing} and \cite{georgiev2019bootstrap} propose the use of a fixed regressor wild bootstrap procedure\footnote{Notice that the particular fixed regressor wild bootstrap procedure is an extension to the fixed regressor bootstrap approach employed by \cite{hansen2000testing} to deal with the appearance of a  nonstandard limiting distribution due to nonstationary regressors.} to approximate critical values. Another related study to the bootstrapping approach of structural break tests is presented by \cite{boldea2019bootstrapping}. However, the proposed framework is restricted to the case of exogenous regressors. 

Despite the recent developments of the structural break literature for predictive regressions, these methodologies consider the detection of parameter instability in predictive regression models without simultaneously testing whether the tests are robust to the presence of predictability around the break-point location. The only existing study that jointly tests for the existence of both effects is the paper of \cite{demetrescu2020testing}, who propose to combine testing for predictability using appropriate subsampling techniques\footnote{The subsampling technique is similarly employed by \cite{davidson2010tests} in the context of detecting structural break due to break in cointegration of the relation under investigation.}(see also similar techniques employed by \cite{hansen2000sample}). The authors propose to use bootstrap-based inference due to the existence of a non-pivotal limiting distribution, which is a robust approach to provide statistical validity of the tests.  

Further aspects related to the IV based approach of KMS that we follow in this paper, can be also examined within the above parameter instability testing procedures.  Recent applications include \cite{magdalinos2020least} who consider predictability tests with GARCH-type effects (see also, \cite{gungor2020small} ) as well as  \cite{yang2020testing} who consider a modification of the KMS test that accounts for serial correlation in the error term of the linear predictive regression. Moreover, \cite{pang2020estimating} propose  testing methodologies for multiple structural breaks under the presence of nonstationary predictors with an application to financial bubble detection. All these features demonstrate additional refinements one can consider as future research related to our proposed framework. 

\newpage 

\paragraph{Outline} The paper is organized as follows. Section \ref{Section2}, presents the predictive regression model along with the background assumptions. This Section also includes a review of the IVX instrumentation procedure of KMS, which is employed for the construction of the proposed test statistic. Section \ref{Section3}, presents the asymptotic theory of the test statistic under different degrees of regressors persistence. Section \ref{Section4} presents an extensive Monte Carlo simulation study. Section \ref{Section5} illustrates an application to the US stock returns which provides evidence of the empirical relevance of the proposed tests for jointly testing predictability and structural break. Section \ref{Section6} concludes. All the mathematical proofs and related results are included in the Appendix of the paper.

\section{Econometric Model and main assumptions}
\label{Section2}

In this section we present the theoretical background within which we operate. We denote with $\left( \Omega, \mathcal{F}, \mathbb{P} \right)$ a suitable probability space on which all of the random elements are defined\footnote{Notice that this is an important assumption, since our framework can be extended to Hilbert spaces within which we can denote the expectation operator and higher moments as inner products. In that scenario, structural break has a slightly different interpretation. We leave this as a future research endeavour. }.  Moreover, throughout the paper, all limits are taken as $T \to \infty$, where $T$ is the sample size. The symbol $"\Rightarrow"$ is used to denote the weak convergence of the associated probability measures as $T \to \infty$. The symbol $\overset{d}{\to}$ denotes convergence in distribution and $\overset{\text{plim}}{\to}$ denotes convergence in probability, within the probability space. Let $\left\{ Y_t, X_t \right\}_{t=1}^T$ denote the corresponding random variables of the underline distributions. 

\subsection{Predictive Regression Model}

Consider the predictive regression model with a possible single structural break
\begin{align}
\label{model1}
y_{t}   &= \alpha_t + \beta_t^{\prime} x_{t-1} + u_{t} \ \ \ \ \ \ \ 0 \leq t \leq T,  
\\
\label{model2}
x_t     &= \mathbf{R}_T x_{t-1} + v_{t}, 
\end{align}
where $y_{t} \in \mathbb{R}$ is an one$-$dimensional vector and $x_t \in \mathbb{R}^{p \times 1}$ is a $p-$dimensional vector of predictors, with an initial condition $x_0 = \mathcal{O}_p(1)$ which we assume that does not affect the limit theory. Moreover,  the autocorrelation matrix is expressed as below
\begin{align}
\label{model3}
\mathbf{R}_T  &= \left( \mathbf{I}_p - \frac{ \mathbf{C} }{ T^{ \gamma_x } } \right) 
\end{align}
where $\mathbf{C} = diag \{ c_1,...,c_p \}$ is a $p \times p$ matrix. The degree of persistence in the regressors is determined by the unknown persistence coefficients $c_i$'s which are assumed to be positive constants and the exponent rate $\gamma_x \in \mathbb{R}$, that is, $\gamma_x < 1$, $\gamma_x \in (0,1)$ or $\gamma_x > 1$. 

\newpage 

The predictive regression model 
given by \eqref{model1}-\eqref{model2} accommodates the existence of a single unknown structural break at location $k$. Therefore, the model parameters take the following form which indicate a time-varying parameter vector (i.e., parameter instability)
\begin{align}
\alpha_t = 
\begin{cases}
\alpha_1 , & 0  \leq t \leq k 
\\
\alpha_1 , & k \leq t \leq T
\end{cases}
\ \ \ \ \text{and} \ \ \ \ 
\beta_t = 
\begin{cases}
\beta_1 , & 0  \leq t \leq k 
\\
\beta_2 , & k \leq t \leq T
\end{cases}
\end{align}
where $\beta_1 \in \mathbb{R}^{p \times 1}$ and $\beta_2 \in \mathbb{R}^{p \times 1}$ while for the univariate model $\beta_1 \in \mathbb{R}$ and $\beta_2 \in \mathbb{R}$. 

To develop  asymptotic theory suitable for robust inference in the predictive regression model under the presence of parameter instability and nonstationarity we impose the following assumptions and regulatory conditions as described below.   
\begin{assumption}
\label{Assumption1}
Let $e_t = \left( u_{t}, v_{t}^{\prime} \right)^{\prime}$ be a $(p+1)-$dimensional vector. The innovation sequence $e_t$ is a conditionally homoscedastic \textit{martingale difference sequence ( m.d.s )}  such that the following two moment conditions hold:   
\begin{enumerate}
\item[A1.] $\mathbb{E} \left[ e_{t} | \mathcal{F}_{t-1} \right] = 0$, where $\mathcal{F}_t = \sigma \left( u_t, u_{t-1},... \right)$ is the natural filtration.
\item[A2.] $\mathbb{E} \left[ e_{t} e_{t}^{\prime} | \mathcal{F}_{t-1} \right] = \mathbf{\Sigma}_{ee}$, where $\mathbf{\Sigma}_{ee} \in \mathbb{R}^{(p+1) \times (p+1)}$ is a positive-definite covariance matrix, which has the following form: 
\begin{align*}
\mathbf{\Sigma}_{ee} =  
\begin{bmatrix}
\sigma_u^2 &  \sigma^{\prime}_{uv} \\
\sigma_{vu} & \mathbf{\Sigma}_{vv}
\end{bmatrix} > 0.
\end{align*}
with $\sigma_u^2 \in \mathbb{R}$, $\sigma_{uv} \in \mathbb{R}^{p \times 1}$ and $\mathbf{\Sigma}_{vv} \in \mathbb{R}^{p \times p}$, where $p$ is the number of predictors.
\end{enumerate}
\end{assumption}
Assumption A1 indicates that the innovation vector is a martingale difference sequence while A2 indicates that the innovation sequence is conditionally homoscedastic. Under conditions A1 and A2, the following Functional Central Limit Theorem (\textit{FCLT}) applies 
\begin{align*}
\frac{1}{ \sqrt{T} } \sum_{t=1}^{ \floor{ T r} } e_t = \frac{1}{ \sqrt{T} } \sum_{t=1}^{ \floor{ T r} } 
\begin{bmatrix}
u_{t} \\
v_{t}
\end{bmatrix}
\Rightarrow \ \text{BM} \left(  \mathbf{\Sigma}_{ee} \right) =
\begin{bmatrix}
B_{u}(r) \\
B_{v}(r)
\end{bmatrix}
= \mathbf{\Sigma}_{ee}^{1/2} W(r),
\end{align*} 
where $\mathbf{\Sigma}_{ee}^{1/2} W(r)$ is a $(p+1)-$dimensional Brownian motion with covariance matrix $\mathbf{\Sigma}_{ee}$. 

\medskip

The above error structure provides a realistic interpretation of macroeconomic shocks. In other words, the shocks to $y_{t}$ and $x_{t-1}$, that is, $u_{t}$ and $v_{t}$ respectively, appear to be contemporaneously correlated, a commonly used assumption in predictive regression models. Related definitions can be found in \cite{ phillips1986multiple}, \cite{phillips1987time} and \cite{phillips1988regression}. Moreover, the implementation of a \textit{FCLT}\footnote{A standard \textit{FCLT} for linear regression models is introduced by Theorem 7.17 in  \cite{white2001asymptotic}.} allows to derive the limiting distribution of Wald type statistics for jointly testing predictability and structural break within the framework of this paper. 

\newpage


\begin{assumption} 
\label{Assumption2}
The innovation to $x_t$ is a linear process with the following representation 
\begin{align*}
v_t 
:= 
\Phi(L) \epsilon_t 
\equiv
\displaystyle \sum_{j=0}^{\infty} \Phi_j \epsilon_{t-j}, \ \ \ \epsilon_t \sim^{i.i.d} \left( 0, \mathbf{\Sigma}_{ee} \right)
\end{align*}
where $\left\{ \Phi \right\}_{j=0}^{\infty}$ is a sequence of absolute summable constant matrices such that $\sum_{j=0}^{ \infty} \Phi_j$ has full rank and $\Phi_0 = I_p$ with  $\Phi (1) \neq 0$, allowing for the presence of serial correlation. 
\end{assumption}
To allow for the detection of possible structural breaks via the predictive regression model and avoid the presence of nuisance parameters at the boundary of the parameter space we also impose the following assumption.
\begin{assumption}
\label{Assumption3}
Let $k = \pi T$, where $0 < \pi < 1$. Then, the fraction of structural break defined as $\pi_0 = k_0 / T$ is within the interior of $(0,1)$ for some fixed $\pi_0$ parameter\footnote{Notice that this mathematical statement does not imply that we consider only known break-points.}.
\end{assumption}
\begin{assumption}
\label{Assumption4}
The estimator $\widehat{\pi}$, is $\mathcal{F}_t-$measurable and $[0,1]-$valued. Then, consistent estimation of the break-point implies that $\exists$ some $m_0$ such that $\widehat{\pi} - \pi = \mathcal{O}_p \left( T^{-m_0} \right)$.
\end{assumption}
\begin{remark}
Assumption \ref{Assumption2} gives the linear process representation of innovations proposed by \cite{phillips1992asymptotics} and can accommodate  the case of serial correlation or even conditional heteroscedasticity with suitable specifications. Assumption \ref{Assumption3} excludes structural breaks at the boundaries of the sample. Assumption \ref{Assumption4} ensures the existence of a consistent estimator of the unknown break-point which imply that $\underset{ T \to \infty }{ \text{lim} } \mathbb{E} \big[ | \hat{\pi} - \pi | \big] = 0$. For instance, assuming that we are testing for an unknown break-point, to avoid unidentified parameters under the null hypothesis the supremum functional is employed for the construction of the Wald type statistics we propose in the next section. 
\end{remark}
When the time-varying parameters $\alpha_t$ and $\beta_t$ are both constant over the full sample, that is, $\alpha_1 = \alpha_2 = \alpha$ and $\beta_1 = \beta_2 = \beta$, then the regression specification given by \eqref{model1}-\eqref{model2} reduces to the standard predictive regression model. In that case, the hypothesis testing of interest is finding statistical evidence against the null hypothesis of no predictability\footnote{For instance, the nonparametric predictive test of \cite{kasparis2015nonparametric} is constructed by comparing the nonparametric functional form estimator with the parametric counterpart estimator. In our setup, the estimation procedure is simpler to that aspect, however it still requires to carefully examine the stochastic properties of the Wald type statistics under nonstationarity and the presence of a single structural break.}. Furthermore, it has been shown in the literature when the ordinary least squares estimation is employed, then statistical inference is nonstandard due to the fact that the limiting distribution of the tests depends on the unknown persistence parameter, $c_i$, and results to uncorrectable asymptotic bias. Since in our setup we are interested in detecting a single structural break an appropriate bias correction which occurs due to endogeneity should take into consideration the presence of these two regimes\footnote{Notice that the underline regimes of our framework are motivated by the hypothesis of parameter instability in the predictive regression model rather than the presence of an independent threshold variable which induces the presence of the two regimes, as in \cite{gonzalo2012regime, gonzalo2017inferring}.}.


\newpage 

\subsection{Robust Inference with the IVX filtration}
\label{Section.sup.Wald.IVX}

To deal with the aforementioned challenges for inference in predictive regression models, a robust procedure has been introduced by \cite{phillipsmagdal2009econometric} and extended by \cite{kostakis2015Robust} to detect the presence of predictability via the IVX-Wald test.  The particular methodology, called IVX filtation, allows for the endogenous formulation of instruments based on the information contained in the regressors of the predictive regression. As a result, the degree of persistence of the instrumental variable has degree of persistence explicitly controlled so that the process is mildly integrated.

Specifically, the IVX filtration considers the first order difference of the corresponding autocorrelation regression such as it can be expressed as below 
\begin{align}
\label{expression.delta}
\Delta x_{t} = \frac{ \mathbf{C} }{ T^{\gamma_x} } x_{t-1} + v_t, \ \ \ \gamma_x \in (0,1).
\end{align}
The particular first difference is not an innovation process unless the regressor belongs to the persistence class of integrated processes. However, it behaves asymptotically as an innovation after linear filtering by a matrix consisting of near-stationary roots\footnote{Note this assumption is a key idea in the development of the asymptotic theory in cointegrated systems for regressors with various types of nonstationarity (see, \cite{phillipsmagdal2009econometric}).}. 

Therefore, the procedure requires to choose an artificial coefficient matrix of the form
\begin{align}
\mathbf{R}_{Tz} =  \left( \mathbf{I}_p - \frac{ \mathbf{C}_z }{ T^{\delta_z} } \right), \ \delta_z \in (0,1), c_z > 0.
\end{align} 
where $\mathbf{C}_z = diag \{ c_{z,1},...,c_{z,p} \}, c_z > 0$ for all $i \in \{1,...,p \}$.  
Then, the instrumental regressor matrix $\widetilde{ z }_t \in \mathbb{R}^{T \times p}$ can be constructed as below
\begin{align}
\widetilde{z}_t = \sum_{j=1}^t \mathbf{R}_{z}^{t-j} \Delta x_j, \ \ \ \mathbf{R}_{z} = \left( \mathbf{I}_p - \frac{ \mathbf{C}_z}{ T^{\delta_z} } \right), \delta_z \in (0,1), \mathbf{C}_z > 0.
\end{align}     
To see this, notice that using expression \eqref{expression.delta} we obtain the following decomposition 
\begin{align}
\widetilde{ z }_t &= \sum_{j=1}^t \mathbf{R}_{z}^{t-j} \left(  \frac{ \mathbf{C} }{ T^{\gamma_x} } x_{j-1} + u_j \right) = \sum_{j=1}^t \mathbf{R}_{z}^{t-j} u_j  + \frac{\mathbf{C}}{ T^{\gamma_x} } \sum_{j=1}^t \mathbf{R}_{z}^{t-j} x_{j-1}.  
\end{align}
which can be written via the following expression  
\begin{align}
\widetilde{ z }_t = z_t + \frac{ \mathbf{C} }{T} \psi_{t},  \ \ \ \ z_t =  \sum_{j=1}^t \mathbf{R}_{z}^{t-j} v_{j} \ \ \text{and} \ \ \psi_{t} = \sum_{j=1}^t \mathbf{R}_{z}^{t-j} x_{j-1}.
\end{align}
Therefore, the IVX filtration proposes to use the constructed $\widetilde{z }_t$ instruments for the regressors $x_t$ which are considered to behave asymptotically as mildly-integrated processes. 

\newpage 

More explicitly, by replacing $x_t$ with the instrument $z_t$ which has a controllable degree of persistence, result to a robust inference procedure which accounts for the effects of nonstationarity. Then, the tuning parameters, which are the exponent rate $\delta_z$ and the diagonal matrix $C_z$ are selected to ensure that $z_t$ is mildly integrated; less persistence than a unit root or a regressor assumed to be generated via a local-unit-root process.   

Thus, the IV based approach, implies that the IVX estimator is expressed as below
\begin{align}
\widetilde{\beta}^{ IVX } = \left[ \sum_{t=1}^T \left( x_t - \bar{x}_{T-1} \right) \widetilde{z}_{t-1}^{\prime} \right]^{-1} \sum_{t=1}^T \big( y_t - \bar{y}_{T} \big) \widetilde{z}_{t-1}^{\prime}, 
\end{align}
where $\bar{x}_{T-1} = \frac{1}{T} \sum_{t=1}^{T} x_{t-1}$ and $\bar{y}_{T} =                      \frac{1}{T} \sum_{t=1}^{T} y_t$, are the corresponding sample means.   

As shown by Theorem A in the Appendix of KMS, the IVX estimator converges to a mixed Gaussian\footnote{The mixed Gaussianity property of the IVX estimator is also extensively examined in the papers of \cite{phillips2013predictive} and \cite{phillips2016robust} under various integration orders.} limiting distribution, which holds regardless of the degree of persistence of the regressors in the model. In turn, this property allows to construct a self-normalized Wald-type statistic which is shown to converge to a standard $\chi^2-$distribution. 

The classical testing hypothesis implies that, $\mathbb{H}_0: \mathcal{R} \beta = 0$ vs $\mathbb{H}_1: \mathcal{R} \beta \neq 0$, where $\mathcal{R}$ the full rank $q \times p$ restriction matrix with rank $q$. Then, the IVX-Wald statistic can be used to test the null hypothesis of no predictability in predictive regression models.  

The IVX-Wald statistic is expressed as below
\begin{align}
\mathcal{W}^{IVX}_T = \widetilde{\beta}^{ IVX \prime } \mathbf{Q}_{\mathcal{R}}^{-1} \widetilde{\beta}^{ IVX}
\end{align}
where $\mathbf{Q}_{\mathcal{R}}$ is a consistent estimator of the asymptotic variance-covariance matrix of $\widetilde{\beta}^{ IVX}$ that accommodates both long-run endogeneity caused by the correlation between the error terms of the system, that is, $u_{t}$ and $v_{t}$, and the finite-sample distortion which results from removing the model intercept. The covariance matrix $\mathbf{Q}_{\mathcal{R}}$, is derived via the following fully modified (FM) estimation of the system covariance terms 
\begin{align}
\mathcal{ \mathbf{Q} }_{\mathcal{R}} 
&= \left( \sum_{t=1}^T \widetilde{z}_{t-1} x_{t-1}^{\prime} \right)^{-1} \mathbf{M} \left( \sum_{t=1}^T x_{t-1}^{\prime} \widetilde{z}_{t-1} \right)^{-1} 
\\
\mathbf{M} &= \widehat{ \sigma }^2_u  \left( \sum_{t=1}^T \widetilde{z}_{t-1} \widetilde{z}_{t-1}^{\prime} \right) - T \bar{z}_{T-1} \bar{z}_{T-1}^{\prime} \widehat{ \mathbf{\Omega} }_{FM} 
\\
\widehat{ \mathbf{\Omega} }_{FM} &=  \widehat{ \mathbf{\Sigma} }_{uu} - \widehat{ \mathbf{ \Omega} }_{uv} \widehat{ \mathbf{\Omega} }_{uu}  \widehat{ \mathbf{\Omega} }_{uv}^{\prime}
\end{align}
where $\widetilde{z}_{T-1} = \frac{1}{T} \sum_{t=1}^{T} \tilde{z}_{t-1}$. Notice that for the univariate predictive regression then $\widehat{ \mathbf{\Omega} }_{FM} \equiv \sigma^2_{FM}$ and $\widehat{\mathbf{\Sigma} }_{uu} \equiv \widehat{ \sigma }^2_u$, where  $\widehat{ \sigma }^2_u$ is a consistent estimator of $\sigma^2_u$. In that case, we can use the bias correction of KMS for the univariate model which is given by 

\newpage 

\begin{align}
M = \left[ \sum_{t=1}^T \widetilde{z}^2_{t-1} - T \bar{z}_{t-1}^2 \left( 1 - \widehat{ \rho }^2_{uv}  \right) \right] \widehat{ \sigma }^2_{u}, \ \ \ \ \ \text{with} \ \ \ \widehat{\rho}^2_{uv} = \displaystyle \frac{ \widehat{\Omega}_{uv} }{ \widehat{\Sigma}_{uu} \widehat{\Omega}_{vv} }.
\end{align}
A key aspect for the computation of the IVX-Wald statistic is the estimation procedure for the matrices $\widehat{ \mathbf{ \Omega} }_{uv}$ and $\widehat{ \mathbf{ \Omega} }_{uu}$, which represent the estimated long-run covariance between $u_{t}$ and $v_{t}$ and the long-run variance of $u_{t}$ respectively. More specifically, these covariance matrices can be constructed using nonparametric kernels with preselected bandwidth parameters, such as the Newey-West type estimators (see, KMS for detailed description and related studies such as \cite{newey1987simple} and \cite{andrews1991heteroskedasticity}).  

Furthermore, it can be proved that under the null hypothesis which imposes a set of linear restrictions in the parameter vector $\beta$, we obtain that $\mathcal{W}^{IVX}_T  \implies \chi^2(q)$ as $T \to \infty$ (see, Theorem 1 in KMS). This important limit theory result provides a unified framework for robust inference and testing regardless the persistence properties of regressors. A simple example is the application of the IVX-Wald test for inferring the individual statistical significance of predictors under abstract degree of persistence. For additional scenarios such as predictors of mixed integration order see \cite{phillips2013predictive} and \cite{phillips2016robust}. In these cases the powerful property of the IVX-Wald statistic still holds.

\subsection{Comparing OLS against IVX based estimators}

In this section, we present some preliminary comparisons between the OLS-Wald and the IVX-Wald statistics, under the null hypothesis of no structural break, to motivate further our research. These comparisons aim to shed some light on the ability of the OLS estimator to detect parameter instability under the assumption of nonstationary predictors. Furthermore, we examine the performance of the IVX estimator for robustifying inference in predictive regression models with persistent\footnote{Notice that the assumption of persistent or integrated regressors is commonly used in the time series econometrics literature due to the nature of economic and financial variables.} predictors under the presence of parameter instability which has been recently a stylized fact in various economic data. 

We consider that the pair $\left\{ \left( y_t, x_t \right): 1 \leq t \leq T \right\}$ is generated by the following bivariate predictive regression system drawn from an i.i.d normally distributed error sequence 
\begin{align}
\label{mod1}
y_t &= \beta_t^{\prime} x_{t-1} + u_t, 
\\
\label{mod2}
x_t &= \left( 1 - \frac{c}{T} \right) x_{t-1} + v_t, \ \ x_0 = 0.   
\end{align}
where $\left( u_t, v_t \right) \overset{ i.i.d }{ \sim } \mathcal{N} \begin{pmatrix}
\sigma_u^2 & \rho \sigma_u \sigma_v  \\
\rho \sigma_u \sigma_v  & \sigma_v^2
\end{pmatrix}$ with $c_i > 0$ or $c_i < 0$ where $y_t \in \mathbb{R}$ and $x_t \in \mathbb{R}$. 

 
\newpage 

We are particularly interested to design Wald statistics robust to nonstationarity which allows to test jointly for the presence of predictability and parameter instability in the predictive regression model given by \eqref{mod1} and \eqref{mod2}. To demonstrate that the estimator of the model parameter can affect the limiting distribution of the test and thus statistical inference we consider via a Monte Carlo experiment testing for a structural break. Thus, we can apply the sup-Wald statistics using each estimator separately by partitioning the sample, estimating the test statistic for each partition and then taking the supremum over the generated sequence of test statistics\footnote{Formal assumptions and asymptotic theory regarding the implementation of the sup-Wald test can be found in \cite{andrews1993tests}. In this section, we demonstrate simple comparisons of the finite sample performance of the two Wald-type statistics based on the OLS and IVX estimators respectively.}.  

Initially, we consider that the covariance matrix $\Sigma$ which captures the dependence between the predictive regression and the persistent regressor has a unit variance. Notice that in this paper we consider a single structural change as in the seminal paper of \cite{andrews1993tests}. When the break-point location, denoted with $\pi$ is known, then to construct the Wald statistic we estimate model parameters $\beta_1 ( \pi )$ for $t = \left\{ 1,..., T \pi \right\}$ and $\beta_2 ( \pi )$ for $t = \left\{ T \pi + 1,..., T \right\}$ based on the observations of the corresponding sub-samples. When the break-point location is unknown then we consider the closed interval $\pi \in [\pi_1, \pi_2]$. 

Under the null hypothesis of no structural break, $\mathbb{H}_0: \beta_1 = \beta_2$, which implies parameter constancy across the full sample. We report rejection frequencies from 5,000 Monte Carlo replications, using the predictive regression model  with a fixed parameter $\beta = 0.25$. For the purpose of comparability we use the critical values that correspond to the NNB asymptotic critical values given by Table 1 of \cite{andrews1993tests}.    The empirical size results for the sup OLS-Wald statistic are presented on Table \ref{Table1} and  \ref{Table2}, while additional results with 10,000 replications can be found on Table \ref{tableC1} and \ref{tableC2}.

As we can see from these rejection frequencies,  under the assumption of persistent predictors (low values of $c$) when we use the critical values that correspond to the standard NBB limiting distribution we obtain size distortions which considerably deteriorate for larger correlation values between the error sequences $u_t$ and $v_t$. Thus, using the OLS estimator especially when modelling nonstationarity can produce biased inference. The econometric intuition is clear, constructing structural break statistics based on the OLS estimator when the predictor follows a local-to-unit root process causes not correctable size distortions of the tests due to the fact that the degree of persistence coefficient $c_i$ is unknown a prior and cannot be consistently estimated (see, Proposition \ref{Proposition1}). Furthermore, when we implement the sup IVX-Wald statistic as explained in the previous section then we will need to carefully examine the stochastic properties of the proposed tests. 

Firstly, we briefly examine in this setting the asymptotic theory of the two estimators to understand which components of the limiting distribution result into nonstandard sampling results especially due to the introduction of parameter instability.  



\newpage   

Notice that the appearance of nonstandard limiting distribution occurs in the case we estimate the sup OLS-Wald statistic, since when we estimate the model parameter $\beta$ by OLS its limiting distribution depends on the following two components  
\begin{align}
\sum_{t=1}^{ \floor{\pi T} } \left( x_{t-1} - \bar{x}_{T-1}  \right)\left( x_{t-1} - \bar{x}_{T-1}  \right)^{\prime} \ \ \ \text{and} \ \ \ \sum_{t=1}^{ \floor{\pi T} } \left( x_{t-1}  - \bar{x}_{T-1}  \right)\left( v_{t-1} - \bar{v}_{T}  \right)^{\prime}.
\end{align}
The first component converges to the integral of a squared demeaned OU process, while the second component weakly converges to a stochastic integral of that demeaned OU process. Therefore, under the assumption of nonstationary regressors and due to the appearance of the corresponding stochastic integral approximations, the second moment of the asymptotic distribution of the model estimator have a nonconstant variance and a nonstandard limiting distribution. For instance, \cite{georgiev2018testing} derive the limiting distribution of the sup-Wald test statistic under the local-to-unit-root generating process, which results to a non-pivotal asymptotic theory. In order to deal with this problem, the authors propose to use the fixed regressor bootstrap which is asymptotically valid even under the nonstationarity assumption. 

Secondly, in the case we estimate the model parameter using the IVX instrumentation, then the following matrix moments appear
\begin{align}
\sum_{t=1}^{ \floor{\pi T} }  \widetilde{z}_{t-1} \left(  x_{t-1} - \bar{x}_{T-1}  \right)^{\prime} \ \ \ \text{and} \ \ \ \sum_{t=1}^{ \floor{\pi T} }  \widetilde{z}_{t-1} \left( v_{t-1} - \bar{v}_{T}  \right)^{\prime}.
\end{align}
The asymptotic theory of the above moment matrices is examined extensively by KMS. A summary of the main results is presented by Lemma \ref{lemma1} in Appendix \ref{AppendixA} of this paper. 

Both OLS and IVX estimators weakly converge to a mixed Gaussian distribution, thus to better understand the differences between the OLS and the IVX estimator\footnote{The asymptotic distribution of the IVX estimator in the univariate predictive regression model is proved to be mixed Gaussian. A proof of this asymptotic result can be found in \cite{phillips2016robust}.} and how these manifest in the asymptotic theory of the tests, we examine the stochastic quantity 
\begin{align}
\mathcal{P}^{IVX}_c \left( \pi \right) :=  \frac{ \displaystyle \pi + \int_0^{\pi} \mathbf{J}_C dJ_C }{ \displaystyle 1 + \int_0^1 \mathbf{J}_C dJ_C}
\end{align}
The $\mathcal{P}_c \left( \pi \right)$ quantity which appears in the limiting distribution of the Wald statistics when an IVX estimator is employed, is a random variable. Thus, considering the analytical form of this random quantity will require to derive an analytical approximation with stochastic terms of higher order, which is a challenging task.

\newpage 

Similarly, the corresponding expression for the OLS estimator is is a function of the stochastic quantity
\begin{align}
\mathcal{P}^{OLS}_c  \left( \pi \right) :=  \frac{ \displaystyle  \int_0^{\pi} \mathbf{J}^2_C dJ_C }{ \displaystyle \int_0^1 \mathbf{J}^2_C dJ_C}. 
\end{align}

To obtain some insights regarding the asymptotic behaviour of the stochastic quantities $\mathcal{P}^{IVX}_c \left( \pi \right)$ and $\mathcal{P}^{OLS}_c  \left( \pi \right)$ we use integral approximation simulation techniques with 10,000 replications for a fixed $\pi \in [ 0.15, 0.85 ]$. Specifically, to approximate the underline OU process which drives these stochastic quantities we use a $1000-$step random walk.  

Moreover, to make these heuristic arguments more apparent, we plot the mean and $95\%$ confidence intervals of the particular empirical fluctuation processes for values of the persistence parameter $c \in \left\{ 1, 5, 100 \right\}$. As we can see from Figure 1, for low persistence levels the asymptotic variance of these processes exhibit a linear growth rate for both the OLS and the IVX estimators since $\mathcal{P}_c  \left( \pi \right)$ has approximate linear increments with $\pi$. In other words,  since the predictive regression model does not assume stationarity for the persistence properties of regressors then by allowing for breaks if we use the OLS based estimator then the accumulation of information as seen by the the second moments has a faster convergence rate especially in the case of high persistence. On the other hand, when we use the IVX estimator this accumulation of information is smoother for different values of $\pi$ due to the properties of the particular estimator in filtering out abstract degrees of persistence, even under time-varying settings.         

FIGURE 1.
\color{black}

\color{black}

\newpage 

\section{Joint Predictability and Structural Break Testing}
\label{Section3}

We focus on the linear predictive regression model and in particular the univariate case. Our proposed testing framework and related asymptotic theory can be extended to the multivariate predictive regression with suitable notation modification. Testing for a single structural break at an unknown break point location within the full sample requires that the multiple predictive regression is expressed as below
\begin{align}
\label{eq.break}
y_{t} &= \left( \alpha_{1} + \beta_{1}^{\prime} x_{t-1} \right) \mathbf{1} \{ t \leq k  \} + \left( \alpha_{2} + \beta_{2}^{\prime}  x_{t-1}  \right) \mathbf{1} \{ t > k \} + u_{t}  
\end{align}
where $k = \floor{ T \pi}$ for some $\pi \in (0,1)$, with $\floor{.}$ denoting the integer part operator. 

To derive the asymptotic theory of the tests we employ the local-unit-root specification proposed by \cite{phillips1987time}. Specifically, since the set of regressors are assumed to be generated via the LUR process $x_t = \left( I_p - \frac{C}{T} \right)$ $x_{t-1} + v_t$, we consider the following $p-$dimensional Gaussian process
\begin{align}
\displaystyle K_c(r) = \int_0^r e^{(r-s)C} d B_v(s), \ \ \ \ r \in (0,1).
\end{align} 
which satisfies the Black-Scholes differential equation
$d K_c(r) \equiv c K_c(r) + d B_v(r)$, with $K_c(r) =0$, implying also that $K_c(r) \equiv \sigma_v J_c(r)$, where $\displaystyle J_c(r) =  \int_0^r e^{(r-s)C} d W_v(s)$ and $K_c(r)$ the Ornstein-Uhlenbeck, (OU) process,\footnote{The OU is a stationary Gaussian process with an autocorrelation function that decays exponentially over time. Moreover, the continuous time OU diffusion process has a unique solution.}
which encompasses the unit root case such that $J_c(r) \equiv B_v(r)$, for $C = 0$. 

\subsection{Classical Least Squares Estimation}

In this Section, we derive the asymptotic theory result that corresponds to a standard sup-Wald test, based on the OLS estimator, when testing for a single structural break in predictive regressions with multiple predictors. 

We denote with $\mathbf{X}_1 := \big[ \mathbf{1} \{ t \leq k  \} \ \ x_{t}  \mathbf{1} \{ t \leq k  \} \big]$, $\mathbf{X}_2 := \big[ \mathbf{1} \{ t > k  \} \ \ x_{t}  \mathbf{1} \{ t > k  \} \big]$ and define $\mathbf{X} = [ \mathbf{X}_1 \ \mathbf{X}_2 ] \in \mathbb{R}^{T \times 2 (p+1)}$ the corresponding partitioned matrix and $\mathbf{\mathcal{R}} = \left[ \mathbf{I}_{p+1} \ - \mathbf{I}_{p+1} \right]$ with $\mathbf{I}_{p+1}$ denoting an identity matrix. Moreover, we denote with $\beta := ( \beta_1  , \beta_2  )^{\prime}$, the parameter vector, then the predictive regression is $y = X \beta + u$. Thus, the OLS Wald statistic for testing the null hypothesis of no structural break, that is, $\mathbb{H}_0: \theta_1 = \theta_2$ against $\mathbb{H}_1: \theta_1 \neq \theta_2$, where  $\theta_j = (\alpha_j, \beta_j )$ for $j=1,2$, is given by the following expression
\begin{align}
\label{Wald.test.ena}
\mathcal{W}_T^{OLS}( \pi ) 
&= \frac{1}{ \hat{\sigma}_u^2 } \left( \widehat{\theta}_1 (\pi) - \widehat{\theta}_2 (\pi)  \right)^{\prime}  \left[ \mathcal{R} \left( X^{\prime} X \right)^{-1} \mathcal{R}^{\prime}\right]^{-1} \left( \widehat{\theta}_1 (\pi) - \widehat{\theta}_2 (\pi) \right) 
\end{align} 

\newpage 

Statistical inference under the null hypothesis of no structural break in the predictive regression are conducted using the supremum functional. The sup OLS-Wald statistic is 
\begin{align}
\widetilde{\mathcal{W}}^{OLS}( t ) := \underset{ \pi \in [ \pi_1 , \pi_2 ] }{ \text{ sup } } \mathcal{W}_T^{OLS}( \pi ),  \ \ \ \text{for} \ \ t_1 \leq  t \leq t_2.
\end{align}
where $ 0 < \pi_1 < 1$ and $ 0 < \pi_2 < 1$ with $\pi_2 = 1 - \pi_1$. 

Under the null hypothesis the breakpoint $k$ is unidentified and so the supremum functional selects the maximum Wald statistic corresponding to a sequence of Wald statistics evaluated at values within the interval $[\pi_1, \pi_2]$. Specifically, in order to construct the corresponding supremum Wald tests we split the full sample $\left\{ y_j, x_{j-1} \right\}_{j=1}^T$ into two sub-samples; the first sub-sample corresponds to the time period before time $t$, $\left\{ y_j, x_{j-1} \right\}_{j=1}^t$  and the second sub-sample corresponds to the time period after time $t$, $\left\{ y_j, x_{j-1} \right\}_{j=t+1}^T$. Due to the fact that we operate within the Skorokhod topology the sample moments that correspond to these sub-samples weakly converge to the corresponding asymptotic result which are based on the $\mathcal{D}[0,1]$ topology. For instance, when deriving the sample moments we can use the notation $t \in \left[ t_1, t_2 \right]$, where $t_1$ and $t_2$ are the lower and upper bounds for the possible break-point location.  

\medskip 

\begin{theorem}
\label{Theorem1}
If Assumptions \ref{Assumption1} and \ref{Assumption3} hold and $\pi$ denotes the unknown break-point, then the sup OLS-Wald statistic under the null hypothesis $\mathbb{H}_0: \theta_1 = \theta_2$ against $\mathbb{H}_1: \theta_1 \neq \theta_2$, where $\theta_j = (\alpha_j, \beta_j )^{\prime}$ with $j=1,2$ based on the predictive regression model \eqref{model1}-\eqref{model2} with $\gamma_x = 1$\footnote{Note that this parameter restriction for the exponent rate implies that $x_t$ follows a LUR process.} weakly converges to the following limiting distribution 
\begin{align}
\label{expression5B}
\widetilde{\mathcal{W}}^{OLS}( \pi ) \equiv \underset{ \pi \in [ \pi_1 , \pi_2 ] }{ \text{ sup } } \ \mathcal{W}_T^{OLS}( \pi )  \Rightarrow \displaystyle \underset{ \pi \in [ \pi_1 , \pi_2 ] }{ \text{ sup } } \ \bigg\{ \widetilde{ \mathbf{N} } ^{\prime}_c( \pi ) \widetilde{ \mathbf{M} }_c( \pi )^{-1} \widetilde{ \mathbf{N} }_c( \pi ) \bigg\}
\end{align}
where
\begin{align}
\widetilde{ \mathbf{M} }_c( \pi )^{-1} = \widetilde{ \mathbf{G} }_c(\pi) - \widetilde{\mathbf{G} }_c(\pi) \widetilde{\mathbf{G} }_c(1)^{-1} \widetilde{\mathbf{G} }_c(\pi)
\end{align}
and
\begin{align}
\widetilde{ \mathbf{N} }_c( \pi ) = \left\{ \widetilde{\mathbf{G} }_c(\pi)^{-1} \widetilde{H} _c(\pi)  - \bigg[ \widetilde{\mathbf{G}}_c(1) - \widetilde{\mathbf{G}}_c(\pi) \bigg]^{-1} \bigg[ \widetilde{H}_c(1) -\widetilde{H}_c(\pi)  \bigg]  \right\}^{\prime}
\end{align}
such that 
\begin{align}
\widetilde{\mathbf{G} }_c(\pi) := \int_0^{\pi} \widetilde{ K }_c( r ) \widetilde{ K }^{\prime}_c( r ) dr \ \ \text{and} \ \  \widetilde{H}_c( \pi ) := \int_0^{\pi} \widetilde{ K }_c( r ) d B_u(r) 
\end{align}
\end{theorem}
with $\widetilde{\mathbf{G} }_c(\pi) \in \mathbb{R}^{ ( p +1 ) \times ( p +1 )}$ a positive-definite stochastic matrix and $\widetilde{H}_c(\pi) \in \mathbb{R}^{ ( p +1 ) \times 1}$. 

\newpage 


Theorem \ref{Theorem1} provides a novel result in the literature and our first theoretical contribution, demonstrating that the sup OLS-Wald statistic as a structural break test on an unknown location based on the predictive regression model with persistent predictors, does not converge to the NBB result as proposed by \cite{andrews1993tests} in linear regressions. Thus, we show that the asymptotic theory of the test depends on the nuisance parameter of persistence $c_i$. However, when the break-point is known a prior, such as $\pi \equiv \pi_0$, it can be easily proved that the limit theory of the OLS-Wald test converges to a NBB even in the case of persistent predictors. The proof of Theorem \ref{Theorem1} is shown in Appendix \ref{AppendixC}.    


\subsection{IVX based Estimation}

In this section, we examine the IVX-Wald based tests for jointly testing for the presence of predictability and parameter instability based on the predictive regression model. The null hypothesis remains the same as in the IVX-Wald test, but the alternative hypothesis is constructed so that it captures both effects. We examine separately the case of stable model intercept and the case of unstable model intercept. The model estimates of the two sub-samples are used to construct the proposed test statistics. We denote with $\widetilde{\beta}_1^{IVX} \left( t \right)$ and  $\widetilde{\beta}_2^{IVX} \left( t \right)$ the IVX estimates of the two sub-samples; and with $\widetilde{Q}_1 \left( t \right)$ and $\widetilde{Q}_2 \left( t \right)$ the corresponding covariance matrices. The sample estimates are given by
\begin{align}
\widetilde{ \beta }_1^{IVX} (t)
&= \left( \frac{1}{t} \sum_{j=1}^t \widetilde{z}_{1,j-1} x_{1,j-1}^{\prime} \right)^{-1} \left( \frac{1}{t} \sum_{j=1}^t \widetilde{ z }_{1,j-1} y_{j}  \right) 
\\
\widetilde{ \beta }_2^{IVX} (t)
&= \left( \frac{1}{T-t} \sum_{j=t+1}^T \widetilde{ z }_{2,j-1} x_{2,j-1}^{\prime} \right)^{-1} \left( \frac{1}{T-t} \sum_{j=t+1}^T \widetilde{ z }_{2,j-1} y_{j} \right).
\end{align}

\subsubsection{Joint Wald Tests under stable model intercept}

We begin our asymptotic theory analysis by considering the case in which the model intercept $\alpha$ is assumed to be stable, that is, under the null hypothesis there is no structural break in the model intercept.  Then, the testing hypothesis of interest is given by 
\begin{align}
\mathbb{H}_0: \alpha_1 = \alpha_2 \ \ \text{and} \ \ \beta_1 = \beta_2
\end{align}
For evaluating our hypothesis we use the IVX-Wald test which has the following form\footnote{Note that the covariance matrices are computed based on the long-covariance matrices and corresponding FM corrections given by KMS, see also \cite{kasparis2015nonparametric}.  These definitions are employed since we do not rule out the strong assumption of a weakly covariance dependence in the model.}
\begin{align}
\mathcal{W}_{\beta}^{IVX}(t) = \left( \widetilde{\beta}^{ IVX }_1 (t) - \widetilde{\beta}^{ IVX }_2 (t)  \right)^{\prime} \bigg[ \widetilde{\mathbf{Q}}_1(t) + \widetilde{\mathbf{Q}}_2(t) \bigg]^{-1} \left( \widetilde{\beta}^{ IVX }_1 (t) - \widetilde{\beta}^{ IVX }_2 (t)  \right). 
\end{align} 

\newpage 



We denote with $\widetilde{\mathcal{W}}^{IVX}_{\beta}(t) \equiv \underset{ t \in [ t_1, t_2 ] }{ \text{sup}} \mathcal{W}_{\beta}^{IVX}(t)$, the corresponding sup IVX-Wald statistic, since we consider the case of the unknown break-point. The two IVX estimates of $\beta$ and their corresponding asymptotic variances are computed, using the data from each sub-sample separately. Thus, the $\mathcal{W}_{\beta}^{IVX}(t)$ statistic can be thought as a Chow-type statistic for detecting break at time $t$, such that $t_1 \leq t \leq t_2$. Then, due to the unknown nature of the break-point the sup Wald test is simply a sequence of Chow-type statistics within the same probability space. Theorem \ref{Theorem2} presents the limiting distribution of $\widetilde{\mathcal{W}}^{IVX}_{\beta}(t)$, under the null hypothesis of no parameter instability in the predictive regression model.

\medskip

\begin{theorem}
\label{Theorem2}
If Assumptions \ref{Assumption1} and \ref{Assumption3} hold and $\pi$ denotes the unknown break-point, then the sup IVX-Wald statistic under the null hypothesis (with $\alpha$ known to be stable a priori) $\mathbb{H}_0 : \alpha_1 = \alpha_2$ and $\beta_1 = \beta_2$, based on the predictive regression model \eqref{model1}-\eqref{model2} and no restriction on the exponent rate $\gamma_x$, has the following asymptotic behaviour
\begin{align}
\widetilde{\mathcal{W}}^{IVX}_{\beta}(t) \Rightarrow \underset{ \pi \in [ \pi_1, \pi_2 ]}{\text{sup}} \bigg\{ \mathbf{N}( \pi )^{\prime} \mathbf{M}( \pi )^{-1} \mathbf{N}( \pi ) \bigg\}
\end{align}
with $t \in [t_1, t_2]$\footnote{Note that, since we operate within the Skorokhod topology $\mathcal{D} \left( 0, 1 \right)$ we can apply standard weakly convergence arguments.} and $\pi \in [\pi_1, \pi_2]$, where 
\begin{align}
\mathbf{N}( \pi )  &= B_p( \pi ) - \mathbf{R}( \pi  ) B_p( 1 )
\\
\mathbf{M}( \pi )  &= \pi \big( \mathbf{I}_p - \mathbf{R}( \pi )  \big) \big( \mathbf{I}_p - \mathbf{R}( \pi )  \big)^{\prime} + (1 - \pi ) \mathbf{R}( \pi ) \mathbf{R}( \pi )^{\prime} 
\end{align}
such that
\begin{equation}
\mathbf{R} ( \pi ) 
=
\begin{cases}
\displaystyle \left( \pi  \mathbf{\Omega}_{xx} + \int_0^{\pi} \mathbf{\underline{B}} dB^{\prime} \right) \left( \mathbf{\Omega}_{xx} + \int_0^{1} \mathbf{\underline{B}} dB^{\prime} \right)^{-1}  & ,\text{if} \ \gamma_x > 1 
\\
\\
\displaystyle \left( \pi  \mathbf{\Omega}_{xx} +  \int_0^{\pi } \mathbf{\underline{J}}_C dJ_C^{\prime} \right) \left( \mathbf{\Omega}_{xx} + \int_0^{1} \mathbf{\underline{J}}_C dJ_C^{\prime} \right)^{-1}  & ,\text{if} \ \gamma_x = 1 
\\
\\
\displaystyle \pi \mathbf{I}_p  & , \text{if} \ \gamma_x < 1
\end{cases}
\end{equation}
where $B(.)$ is a $p-$dimensional standard Brownian motion, $J_C (\pi) = \int_0^{\pi} e^{C (\pi - s)} dB(r)$ is an \textit{Ornstein-Uhkenbeck} (OU) process and we denote with $\underline{J}_C (\pi) = J_C (\pi) - \int_0^1 J_C(s) ds$ and $\underline{B} (\pi ) = B(\pi) - \int_0^1 B(s) ds$ the demeaned processes of $J(\pi)$ and $B(\pi)$ respectively.
\end{theorem}
Theorem \ref{Theorem2} demonstrates that the supremum functional of the IVX-Wald test weakly convergence to a stochastic quadratic functional and does not have the same asymptotic behaviour as the corresponding IVX-Wald statistic which is proved to follow a $\chi^2_p$ distribution (see, KMS, \cite{phillips2013predictive} and  \cite{phillips2016robust}). 


\newpage 


An important implication of Theorem \ref{Theorem2}, which consists our second contribution to the limit theory of structural break tests in predictive regression models with nonstationary predictors, is that we show that the dependence of the limiting distribution to the unknown parameter of persistence takes different forms by restricting further the parameter space of the exponent rate. Specifically, Theorem \ref{Theorem2} demonstrates that when testing for a structural break in predictive regression models, under the assumption that the predictors are integrated or nearly integrated, then the limiting distribution of the test, weakly converge to a nonstandard process, which verifies the corresponding result already mentioned by \cite{hansen2000testing}). On the other hand, when the predictors are stationary or mildly stationary, then the test statistic weakly converges to the familiar squared tied-down Bessel process as proved by \cite{andrews1993tests} in the case of linear regression models. The latter implies that when we consider separately the special case for which $\gamma_x < 1$, which covers both the cases of mildly integrated regressors, i.e., $\gamma_x \in (0,1)$, and stationary regressors, i.e., $\gamma_x = 0$, then it can be proved that the limiting distribution of the sup IVX-Wald test converges to the standard NBB result\footnote{Note that in the case of a known break-point we can easily prove a convergence to a $\chi^2_p$ distribution which is free of nuisance parameters and so conventional inference methods apply}. Corollary \ref{corollary1} below is a direct implication of Theorem \ref{Theorem2} and summarizes this finding.

\medskip

\begin{corollary}
\label{corollary1}
Under the assumptions and definitions given by Theorem \ref{Theorem2}, when $\gamma_x < 1$ then the following asymptotic distribution holds  
\begin{align}
\widetilde{\mathcal{W}}^{IVX}_{\beta}(t) \Rightarrow \underset{ \pi \in [ \pi_1, \pi_2 ]}{\text{sup}} \frac{ \mathcal{BB}_p ( \pi )^{\prime} \mathcal{BB}_p ( \pi )  }{ \pi (1 -  \pi) }, 
\end{align}
where $\mathcal{BB}_p ( . )$ is a $p-$dimensional standard Brownian bridge. 
\end{corollary} 

\begin{remark}
The weakly convergence of the sup IVX-Wald test into a normalized squared Brownian bridge, when the exponent rate is restricted such that $\gamma_x < 1$, implies standard statistical inference due to the known distribution. One rejects $\mathbb{H}_0$ for large values of the sup IVX-Wald test based on  a significance level $\alpha$ such that $0 \leq \alpha \leq 1$ and thus the limit distribution can be used to derive associated critical values, denoted with $c_{\alpha}$ such that $\mathbb{P} \big( \widetilde{\mathcal{W}}_{\beta}^{IVX} ( \pi ) > c_{\alpha} \big) > 0$ with $\underset{ T \to \infty }{ \text{lim} } \mathbb{P} \big( \widetilde{\mathcal{W}}_{\beta}^{IVX} ( \pi ) > c_{\alpha} \big) = 1$.  
\end{remark}

Our findings presented by Theorem  \ref{Theorem2} verify the conjuncture of \cite{hansen2000testing} who argues that when testing for a structural break based on a sup-functional induces an asymptotic non-pivotal distribution under the assumption of nonstationarity. However, this seminal study has not examined in details certain forms of nonstationarity which can occur and how these are manifested in the limit theory of the tests. In particular, within our framework we demonstrate using local-to-unit root asymptotic arguments that the nonstandard limiting distribution occurs when  $x_t$ is properly modelled as a nonstationary stochastic process, and specifically the NBB result no longer holds when $x_t$ is either a nearly integrated or an integrated process.

\newpage 

The proofs of both Theorem \ref{Theorem2} and Corollary \ref{corollary1} can be found in Appendix \ref{AppendixA}. Next, we focus on designing test statistics for jointly testing against predictability and structural break in predictive regression models. We define a joint Wald test based on the IVX estimator under the assumption of an unknown break-point as below 
\begin{align}
\label{joint.wald.ivx}
\widetilde{ \mathcal{W} }_{\beta}^{joint} := \underset{ t \in [ t_1, t_2 ] }{ \text{sup}  } \bigg\{ \mathcal{W}_{T}^{IVX} + \mathcal{W}_{\beta}^{IVX}(t) \bigg\}, \ \ \ \text{for} \ \ t_1 \leq  t \leq t_2.
\end{align}
where the supremum functional applies only on the second component of the test above. Then, the Corollary \ref{corollary2} below gives the related limit theory result. 
  
\begin{corollary}
\label{corollary2}
\textit{(i)} If the conditions of Theorem \ref{Theorem2} hold, then under the null hypothesis, $\mathbb{H}_0: \alpha_1 = \alpha_2$ and $\beta_1 = \beta_2 = 0$ and no restriction on the exponent rate $\gamma_x$, the large sample theory of the test statistic specified by \eqref{joint.wald.ivx} has the following form 
\begin{align}
\widetilde{ \mathcal{W} }_{\beta}^{joint} \Rightarrow B(1)^{\prime} B(1) + \underset{ \pi \in [ \pi_1, \pi_2 ]}{ \text{sup} } \bigg\{ \mathbf{N}( \pi )^{\prime} \mathbf{M}( \pi )^{-1} \mathbf{N}( \pi ) \bigg\},
\end{align}
where $B(.)$, $\mathbf{N}( \pi)$ and $\mathbf{N}( \pi )$ are defined in Theorem \ref{Theorem2}. 
\\
\textit{(ii)} As a special case, when $\gamma_x < 1$, it follows that 
\begin{align}
\widetilde{\mathcal{W}}_{\beta}^{joint} \Rightarrow \chi^2_p + \underset{ \pi \in [ \pi_1, \pi_2 ]}{ \text{sup} } \frac{ \mathcal{BB}_p ( \pi )^{\prime} \mathcal{BB}_p ( \pi  )  }{ \pi (1 -  \pi ) },
\end{align}
where $\mathcal{BB}_p(.)$ is a $p-$dimensional standard Brownian bridge and $\chi^2_p$ denotes the $\chi^2$ random variable with $p$ degrees of freedom. Furthermore, the two stochastic quantities of the limiting distribution are assumed to be independent.    
\end{corollary}

\begin{remark}
Notice that Corollary \ref{corollary2} demonstrates that when $\gamma_x < 1$, the large sample theory of the test statistic $\mathcal{W}_{ \beta }$ is pivotal. Therefore, in this special case by having a limiting distribution being free of nuisance parameters, asymptotic critical values for testing the null hypothesis can be easily obtained. The particular test statistic provides a methodology for testing for both predictability and structural break which is robust to the persistence properties of regressors after replacing the OLS with an IVX estimator.   
\end{remark}
The main arguments we use for the proof of Corollary \ref{corollary2} is to consider the joint testing hypothesis as a composite hypothesis based on the mutually exclusive parameter space, and thus we construct the limit theory of this test based on the joint formulation of the two separate testing hypotheses. By employing the asymptotic matrix moments based on the IVX instrumentation we prove that the limiting distribution of the joint test can be decomposed into two components. These two components are considered to be independent random variables and therefore practically we could use the critical values of the limiting distributions that corresponds to each of these stochastic quantities. 


\newpage 

\subsubsection{Joint Wald Tests under unstable model intercept}



Notice that when a model intercept is included then under the null hypothesis, when $\alpha_1 = \alpha_2$, for example if we are testing whether $H_0: \beta_1 = \beta_2$, then the particular hypothesis would be equivalent to testing the null hypothesis of no structural break. However, in the case of an unstable model intercept, that is, $\alpha_1 \neq \alpha_2$, then testing the null hypothesis $H_0: \beta_1 = \beta_2$ can result to a different asymptotic theory due to the break in the model intercept. In this section, we consider the case that a potential structural break occurs in both the intercept and the slope coefficient $\beta$ of the predictive regression model.  The two proposed tests aim to test: (i) the null hypothesis of no break on model intercepts while the slope coefficients remain at a fixed level; (ii) the null hypothesis of no break on model intercepts while the slope coefficients are both equal to zero. 

The proposed test statistic is expressed as below
\begin{align}
\mathcal{W}^{IVX}_{\alpha}(t) =  \bigg( \tilde{\alpha}_1 (t) - \tilde{\alpha}_2 (t)  \bigg)^{\prime} \bigg[ \widetilde{\mathbf{\Omega} }_1(t) + \widetilde{ \mathbf{\Omega} }_2(t) \bigg]^{-1} \bigg( \tilde{\alpha}_1 (t) - \tilde{\alpha}_2 (t)  \bigg) 
\end{align}
where the covariance estimators are computed as below
\begin{align}
\label{correction1}
\widetilde{\mathbf{\Omega}}_1 (t) &= \frac{1}{t^2} \widehat{u}_1(t)^{\prime}\widehat{u}_1(t)  + \bar{z}_t^{\prime} \widetilde{\mathbf{Q}}_1(t) \bar{z}_t 
\\   
\label{correction2}
\widetilde{\mathbf{\Omega}}_2 (t) &= \frac{1}{( T - t )^2} \widehat{u}_2(t)^{\prime}\widehat{u}_2(t)  + \bar{z}_{T- t}^{\prime} \widetilde{\mathbf{Q}}_2(t) \bar{z}_{T- t} 
\end{align}
Then, the joint Wald test is expressed using the following form 
\begin{align}
\mathcal{W}_{\alpha \beta }^{joint} :=  \bigg\{  \mathcal{W}_{T}^{IVX}  +  \mathcal{W}^{IVX}_{\alpha}(t) + \mathcal{W}^{IVX}_{\beta}(t)  \bigg\}, \ \ \ \text{for} \ \ t_1 \leq  t \leq t_2.
\end{align}
To distinguish between the standard Wald statistics in the case of known break-point and the Wald type statistics that correspond to the unknown break-point we denote with $\widetilde{\mathcal{W}}^{IVX}_{\alpha}(t) = \underset{ t \in [ t_1, t_2 ]}{\text{sup}} \mathcal{W}^{IVX}_{\alpha}(t)$ and $\widetilde{\mathcal{W}}^{joint}_{\alpha \beta }(t) = \underset{ t \in [ t_1, t_2 ]}{\text{sup}} \mathcal{W}^{joint}_{\alpha \beta}(t)$. Both of these test statistics require to split the sample into subsamples of window size $[ t_1, t_2 ] \subset (1,T)$ and estimate the joint Wald test based on the observations from each subsample. For instance, the last two components of the joint test requires to estimate the Wald IVX for the full sample plus the supremum statistics for the model intercept and the model slope separately. Notice that the model intercept has different asymptotic properties, in particular, convergence rate to the true population parameter when the IVX estimator is used for the slope parameter. Therefore, we use the covariance estimators given by \eqref{correction1} and  \eqref{correction2} which is constructed based on the residuals from the fitted predictive regression of each subsample while we add a bias correction based on the covariance estimator of the slope coefficients since the estimator of the intercept is conditional on the IVX estimator in the case we switch from the OLS to the IVX estimation procedure. 

\newpage 

\begin{proposition}
\label{Proposition1}
Consider the predictive regression model given by expressions \eqref{model1}-\eqref{model2}. If Assumption \ref{Assumption1}-\ref{Assumption4} hold and $\alpha$ is known to be unstable  a priori, then under the null hypothesis $\mathbb{H}_0: \alpha_1 = \alpha_2 \ \ \text{and} \ \ \beta_1 = \beta_2 = \beta$, as T $\to \infty$ the following limit result holds
\begin{align}
\widetilde{\mathcal{W}}^{IVX}_{\alpha}(t) \Rightarrow \underset{ \pi \in [ \pi_1, \pi_2 ]}{ \text{sup} } \frac{ \mathcal{BB}_1 ( \pi )^{\prime} \mathcal{BB}_1 ( \pi )  }{ \pi (1 - \pi) }
\end{align}
where $\mathcal{BB}_1 ( . )$ is a one-dimensional standard Brownian bridge. 
\end{proposition} 

\begin{remark}
Notice that Proposition \ref{Proposition1} provides an asymptotic result for a composite hypothesis since we consider jointly testing for a structural break in the model intercept and the slope coefficients while we test, that under the null hypothesis the slope coefficient has a fixed parameter value $\beta$. Additionally, we can investigate the limiting distribution of the joint Wald test when we assume that under the null hypothesis there is no predictability.      
\end{remark}

\begin{proposition}
\label{Proposition2}
Consider the predictive regression model given by expressions \eqref{model1}-\eqref{model2}. If Assumption \ref{Assumption1}-\ref{Assumption4} hold and $\alpha$ is known to be unstable  a priori, then under the null hypothesis $\mathbb{H}_0: \alpha_1 = \alpha_2$ and $\beta_1 = \beta_2 = 0$, as $T \to \infty$ the following limit results hold: 
\
(i)
\begin{align}
\widetilde{\mathcal{W}}^{joint}_{\alpha \beta }(t) \Rightarrow  B(1)^{\prime} B(1) + \underset{ \pi \in [ \pi_1, \pi_2 ]}{ \text{sup} } \bigg\{ \widetilde{ \mathbf{N} }( \pi )^{\prime}  \widetilde{ \mathbf{M} } ( \pi )^{-1} \widetilde{ \mathbf{N} }( \pi ) \bigg\}
\end{align}
where $\widetilde{ \mathbf{N} }( \pi ) = \big( \mathcal{BB}_1( \pi ), \mathbf{N}( \pi ) \big)^{\prime}$ and $\widetilde{ \mathbf{M} }( \pi ) = 
\begin{pmatrix}
\pi(1- \pi) & 0
\\
0 & \mathbf{M} (\pi)
\end{pmatrix}$. The terms, $\mathbf{N}(\pi)$ and $\mathbf{M} (\pi)$ are defined in Theorem \ref{Theorem2}. 
\\
\textit{(ii)} As a special case, when $\gamma_x \in (0,1)$, it holds that 
\begin{align}
\widetilde{\mathcal{W}}^{joint}_{\alpha \beta }(t) \Rightarrow \chi^2_p + \underset{ \pi \in [\pi_1, \pi_2 ]}{ \text{sup} } \frac{ \mathcal{BB}_{p+1} ( \pi )^{\prime} \mathcal{BB}_{p+1} ( \pi )  }{ \pi (1 -  \pi) }
\end{align}
where $\mathcal{BB}_{p+1} ( . )$ is a $(p+1)-$dimensional standard Brownian bridge, and $\chi^2_p$ is a random variable following a $\chi^2$ distribution with $p$ degrees of freedom.
\end{proposition} 
\begin{remark}
Notice that, Proposition \ref{Proposition2} shows that the limiting distribution of the joint test for both predictability and structural break, when we consider simultaneously testing whether there is a structural break to the model intercept and no predictability using the set of regressors of the model, has an asymptotic distribution which takes a different form when we consider different values of the parameter space of the exponent rate. 
\end{remark}
As we have seen by our extensive asymptotic theory analysis provided in this Section, the asymptotic distribution of the Joint IVX-Wald tests can be affected by various scenarios, such as the inclusion of model intercept in the predictive regression as well as the parameter space of the exponent rate. 





\newpage

\newpage 

\section{Monte-Carlo Simulation Study}
\label{Section4}

In this section, we present a Monte Carlo simulation study in order to examine the finite size properties of the proposed Wald-type statistics in terms of their empirical size and power performance, under the null hypothesis of no joint parameter instability and predictability. In practise, the degree of persistence in the time series of the regressors is unknown. That is, both the coefficient of persistence $c_i$ as well as the exponent rate $\gamma_x$ are both unknown parameters to the researcher. Moreover, we have proved that the limiting distribution of the Wald-type statistics for detecting structural break in predictive regression models depend on these unknown properties of the regressors for certain parameter value restrictions on the coefficient of persistence. In particular when the exponent rate of persistence has an exponential rate , then for both cases i.e., testing for structural break or jointly testing for predictability and structural break we have a nonstandard limiting distribution which depends on the corresponding stochastic integrals which are functions of the unknown break-point. Now, in the case we known a prior that the exponent rate equals to one, which means we are within the realm of near integrated or LUR regressors then the tests converge to a standard NBB distribution which is easy to tabulate critical values. Nevertheless, in any case in empirical applications critical values can be obtained by either simulations or more advanced bootstrap methodologies. 

The MC simulation study aims to shed light on the main theoretical results of the paper. Therefore, to demonstrate the above theoretical result, under the null hypothesis of no structural break, we can generate a DGP with no breaks in the coefficients of the predictive regression. Then, constructing the sup-Wald test and using the Andrews' critical values we can observe whether size distortions indeed occur in this scenario\footnote{Recall that  under the assumption of persistent predictors, we prove via Proposition \ref{Proposition1} of the paper that when testing for parameter instability in the predictive regression the limiting distribution of the sup-Wald statistic no longer follows a normalized squared brownian bridge, (NBB) for an unknown structural break. In particular, when contacting inference or assessing the statistical validity of the sup-Wald statistic via a MC experiment, using the corresponding critical values of the sup-Wald test proposed by \cite{andrews1993tests} we can observe that leads to size distortions due to the non-NBB limiting distribution.}. Secondly, via Proposition \ref{Proposition2} of the paper we propose an alternative approach to overcome this problem. In particular, using an IV based sup-Wald test, which is constructed using the IVX instrumentation, we prove that the limiting distribution of the statistic indeed weakly converges to a NBB, which allow us to use the Andrews' critical values, avoiding this way to simulate critical values which can be computational complex. Below, we present the DGP, the test statistics as well as the size and power comparisons for the proposed tests of our econometric framework.

\newpage 

\subsection{Experiment Design}

We use the following data generating process (DGP) where $y_t$ is a scalar and $x_t$ is a vector of LUR predictors (with the property of being highly persistent).
\begin{align}
\label{m1}
y_t &= \beta_0 + \beta_1 x_{t-1} + \beta_2 x_{t-1} +  u_t \\
\label{m2}
\underline{x}_t &= 
\begin{pmatrix}
1 - \displaystyle \frac{c_1}{T} & 0 \\
0  & 1 - \displaystyle \frac{c_2}{T}
\end{pmatrix}   
\underline{x}_{t-1} + \underline{v}_t,  \ \underline{x}_0 = \underline{0}.
\end{align}
with $t \in \{1,...,T \}$ and $T = \{ 100, 250, 500, 1000 \}$ for $B = 5,000$ replications. 

Furthermore, we consider the effect of different localizing coefficients of persistence across the predictors. We use $c_i \in \{ 1, 5, 10, 20 \}$ for $i = 1,2$, which cover various cases of LUR regressors, with smaller values implying that we impose the assumption of higher persistence and lower values implying that existence of mild persistence in the predictor.  


The covariance matrix of the innovations $\underline{e}_t = \left( u_t, \underline{v}_t  \right)^{\prime} \sim \mathcal{N} \left( \underline{0}_{3 \times 1}, \Sigma_{ee} \right)$ we assume that is parametrised with the following structure 
\begin{align}
\label{Sigma.ee}
\Sigma_{ee}
=
\begin{bmatrix}
\sigma^2_u & \sigma_{uv_1} &  \sigma_{uv_2} \\
\sigma_{uv_1}  & \sigma^2_{v_1} &  \sigma_{v_1 v_2} \\
 \sigma_{uv_2} & \sigma_{v_1 v_2} &   \sigma^2_{v_2}
\end{bmatrix}
\end{align}
Clearly, we can see that the covariance matrix given by expression \eqref{Sigma.ee} allows to consider various scenarios regarding the contemporaneous correlation of the regressors and the dependent variables which has related economic interpretation. We use the following predetermined covariance matrices 
\begin{align*}
\Sigma_{ee} = 
\begin{bmatrix}
1 & 0 & 0 \\
0 & 1 & 0 \\
0 & 0 & 1 \\
\end{bmatrix}, \ \ \ \text{or} \ \ \ 
\Sigma_{ee} = 
\begin{bmatrix}
1     & 0.10   & -0.29 \\
0.10  & 1      & -0.03 \\
-0.29 & -0.03  & 1    \\
\end{bmatrix} 
\end{align*}

Under both the null and the alternative hypothesis, the predictive regression is generated using the following econometric specification (using \eqref{m2} and \eqref{Sigma.ee})
\begin{align}
y_{t} &= \left( \beta_{01} + \underline{\beta}_{11}^{\prime} x_{t-1} \right)  \mathbf{1} \{ t \leq k  \} + \left( \beta_{02} + \underline{\beta}_{12}^{\prime} x_{t-1} \right) \mathbf{1} \{ t > k \} + u_{t}  
\end{align}
where $k = \floor{ T \pi}$ for some $\pi \in (0,1)$. 

Additionally under the null hypothesis, of no parameter instability, we simulate the DGP using the same vector of regression parameters before and after the breakpoint, such as $\underline{\beta}_1 = \underline{\beta}_2$. Furthermore, under the alternative hypothesis of parameter instability, we can generate a sequence of local alternatives by using a different vector of regression parameters before and after the breakpoint. Note that, since our framework consider a single unknown structural break, the DGP is re-constructed for each window $[\pi_1, \pi_2]$ in order to apply the supremum functional. 

\subsection{Test statistics}
\label{test.statistics}

We assess the statistical validity in finite and large samples of the Wald based statistics within our proposed framework, that is, the sup Wald-OLS test and the sup-Wald IVX test. In particular, we are interested to verify any size distortions under the null hypothesis of no structural change, when using the sup Wald-OLS test and we expect to observe improvements in the empirical size of the sup Wald-IVX test. Moreover, we examine the size and power performance of the two statistics across different degree of persistence as well as the rate at which we allow the IVX instrumentation procedure to create a more mildly integrated regressor (as given by the parameters $c_z$ and $\delta \equiv \delta_0$). 


For the large sample properties of the test statistics a standard convergence result apply, that is, $\mathcal{W}_T ( \pi ; \delta ) \to \mathcal{W} ( \pi ; \delta )$ as $T \to \infty$. Furthermore, the testing hypothesis of interest is a two-sided type hypothesis which is expressed as below 
\begin{align}
\mathbb{H}_0: \underline{\beta}_1 = \underline{\beta}_2 \ \ \ \text{vesus} \ \ \  \mathbb{H}_1: \underline{\beta}_1 \neq \underline{\beta}_2 
\end{align}
The test statistics are computed via expressions \eqref{wald.ols} and \eqref{wald.ivx}, while standard regularity and invariance principles holds (e.g, uniform convergence for the compact space $\underline{\theta} \in \Theta$ such that $( \underline{\beta}_1, \underline{\beta}_2)^{\prime} \subset \underline{\theta}$ and $\Theta \in \mathbb{R}^q$).  

\subsubsection{Wald-OLS statistic}

\begin{align}
\label{wald.ols}
\mathcal{W}^{\text{OLS}}_T( \pi )
&= 
\frac{1}{ \hat{\sigma}_u^2 } \left( \hat{ \underline{\beta} }_{1} - \hat{ \underline{\beta} }_{2} \right)^{\prime}  \left[ \left(  X_1^{\prime} X_1 \right)^{-1} + \left( X_2^{\prime} X_2 \right)^{-1} \right]^{-1} \left( \hat{ \underline{\beta} }_{1} - \hat{ \underline{\beta} }_{2} \right) 
\end{align}

\subsubsection{Wald-IVX statistic}

\begin{align}
\label{wald.ivx}
\mathcal{W}^{\text{IVX}}_T( \pi ) 
&= 
\frac{1}{ \hat{\sigma}_u^2 } \left( \hat{ \underline{\beta} }^{\text{IVX}}_{1} - \hat{ \underline{\beta} }^{\text{IVX}}_{2} \right)^{\prime}  \mathcal{Q}_{ \mathcal{R} }^{-1} \left( \hat{ \underline{\beta} }^{\text{IVX}}_{1} - \hat{ \underline{\beta} }^{\text{IVX}}_{2} \right) 
\end{align}
where $\mathcal{Q}_{ \mathcal{R} }$ is defined as below
\begin{align}
\label{cov.matrix1}
\mathcal{Q}_{ \mathcal{R} }  = \left\{  \left(Z_1^{\prime} X_1 \right)^{-1} \left( Z_1^{\prime} Z_1 \right) \left(X_1^{\prime} Z_1 \right)^{-1}  
+  
\left(Z_2^{\prime} X_2 \right)^{-1} \left( Z_2^{\prime} Z_2 \right) \left(X_2^{\prime} Z_2 \right)^{-1}  \right\} 
\end{align}

In the case of a known break-point we can use the critical values proposed by \cite{andrews1993tests} with an appopriate significance size such as $\alpha = 5\%$ for both test statistics. Since we consider an unknown break point, then the sup functional gives the statistic with the maximum value after estimating a sequence of statistics over the interval $\pi \in [ \pi_1, 1 - \pi_2]$, and thus critical values need to be used due to the fact that the limiting distribution in this case is not a standard $\chi^2-$distribution. Furthermore, the data that we generate from the DGP given by \eqref{m1} and \eqref{m2} only differ over the values of $c_1$ and $c_2$ and are the same for different values of the sample size $T$. Thus, as the sample size increases the degree of persistence of the endogenous regressors remain the same one (for both predictors). Moreover, the increase in the sample size aims to reflect the properties of finite versus large sample asymptotics for the Wald type statistics in testing for a single unknown break-point in predictive regression models with persistent regressors. Furthermore, in order to control for the existence of empirical size distortions when testing for structural break via the Wald-IVX statistic, we use the bias corrected IVX estimators as proposed by \cite{phillipsmagdal2009econometric}. The particular bias correction, implies that we account for an overestimation effect when projecting the instrumental variable towards the direction of the dependent variable (due to endogeneity). 


The bias corrected IVX estimator has the following form 
\begin{align}
\tilde{ \underline{\beta} }^{\text{IVX}} = \left( Y^{\prime} \widetilde{Z} - T \hat{\Delta}_T (h,m )  \right) \left( X^{\prime} \widetilde{Z} \right)^{-1}
\end{align}  
where the estimator of $\Delta$ is given by
\begin{align}
\hat{\Delta}_T (h,m ) = \frac{1}{T} \sum_{h=0}^m \left( 1 - \frac{h}{m+1} \right) \sum_{t = h+1}^T  \hat{u}_t \hat{u}^{\prime}_{t-h} 
\end{align}
The above non-parametric estimator is known as Newey-West type estimator (see, \cite{newey1987simple}) and allows to estimate the bias correction without imposing additional parametric assumptions. Optimal bandwidth choices can be of the form $m = \eta T^{1/5}$, where $\eta$, is a positive constant. The bias correction is applied to both $\tilde{ \underline{\beta}_1 }^{\text{IVX}}$ and $\tilde{ \underline{\beta}_2 }^{\text{IVX}}$.

\subsection{Size Comparison}

To conduct a size comparison of the test statistics, we examine the empirical rejection rates of the sup-Wald OLS and sup-Wald IVX tests for detecting single structural change with persistent predictors, under the null hypothesis of no structural change, that is, $\mathbb{H}_0: \underline{\beta}_1 = \underline{\beta}_2$ using a significance level $\alpha = 5\%$ and compare using different critical value approximations. In particular, we repeat the empirical-size experiment using the asymptotic critical values from (i) Table 4 of \cite{gonzalo2012regime}, (ii)  Table 1 of \cite{andrews1993tests}), (iii) the bootstrap generated critical values within the MC step. More specifically, to do this, we generate 5,000 datasets from DGP \eqref{m1} and \eqref{m2} for various values of $\underline{\beta}^0$ and compute the frequency of rejecting the null hypothesis.
\begin{align*}
\textbf{Design 1} \ \ \ \ \ \ \ \ \ \underline{\beta}^0 &:= \big( \beta_0 = 0.25, \beta_1 = 0.50  \big)
\\
\\
\textbf{Design 2} \ \ \ \ \ \ \ \ \ \underline{\beta}^0 &:= \big( \beta_0 = 1, \beta_1 = 0.8, \beta_2 = 1.2   \big)
\end{align*}

Table \ref{Table1} (for sup Wald-OLS) and Table \ref{Table2} and \ref{Table3} (for sup Wald-IVX), present the probabilities of rejection of the two Wald-type statistics at the $5\%$ nominal rate, under the null hypothesis (for Design 1). In particular, we consider different values for the exponent rate of the IVX parameter, such as $\delta \in \{ 0.75, 0.95 \}$ in order to investigate the varying effect of the degree of persistent of the instrumental variable for detecting structural change in predictive regressions with persistent regressors, as well as different localizing coefficient of persistence, $c_i \in \{ 1, 5, 10, 20 \}$. Comparing the empirical size results for Table \ref{Table1} versus Table \ref{Table2} and \ref{Table3}, we can see that the sup Wald IVX produces values for the empirical size quite close to the nominal size for critical value $c_{\alpha} = 13.42$, $\delta = 0.95$ and $ -0.5 \leq \rho \leq 0.5$. The particular critical value is a closer representation to the $\alpha-$quantile from the corresponding limiting distribution. 

In Table \ref{Table1} and Table \ref{Table2} we present the empirical size under the null hypothesis for the model with a single predictor and no-intercept. We can observe that size distortions appear for larger values of correlation between the $u_t$ and $v_t$ and this is more severe for high persistence regressors (i.e., low values of the coefficient of persistence). Moreover, we also consider the case of explosive regressors, that is, $c_i < 0$. Clearly, the empirical size in this scenario has higher values since we use the critical value of Andrews (8.85) which is not the corresponding critical value of the asymptotic distribution of the test statistic for detecting structural break in the predictive regression model under the assumption of explosive regressors. Notice also that in the case we have $p > 1$ and we use the sup OLS-Wald statistic for testing for joint predictability and structural break then  the size distortions are much higher than in the case of $p =1$ and this is due to the nonstandard limiting distribution of the test statistic under the assumption of nonstationary regressors.  

All main conclusions are in line with similar findings in the literature of predictive regression models, such as larger size distortions appear as the correlation between u(t) and v(t) increases, and this effect is more apparent in the case of persistent predictors (i.e., lower values of the persistence coefficient c). Moreover, we see that the Sup OLS-Wald statistic with the standard asymptotics (Andrews) is clearly immune to persistence when there is no correlation between u(t) and v(t). Furthermore, size distortions appear under the existence of nonzero correlation between the error sequences $u_t$ and $v_t$. 

\color{black}

\newpage 

In other words, when Cov $\left( u_t, v_t \right) \neq 0$, then under the assumption of nonstationary predictors which is captured via the Local to unit root specification, then the limiting distribution no longer follows the standard NBB result of Andrews. In fact in this case it depends on the degree of persistence $c_i$. 

We observe that even though the simulated asymptotic values of the sup IVX-Wald test for detecting a single structural break in the predictive regression model with no model intercept is not quite close to the 8.85 cv that corresponds to the NBB result of Andrews, we can verify that the asymptotic result does not depend on the nuisance parameter $c$.   

Moreover, from the empirical simulations across different values of $\rho$ we can see that the simulated asymptotic critical values are quite close when observing at a specific value of $c$. This is not surprising since the fully modified covariance estimator incorporated in the construction of the covariance matrix for the IVX takes into account the dependence structure of the regressors by applying a common long-run covariance structure. This holds across different values of $c_i$.  
\color{black}

\subsection{Illustrative Examples}

We consider for instance the case of bivariate predictor persistence and examine the finite-sample performance of the proposed tests. In this case, we use the $c_1 = 1$ and $c_2 = 5$ considering this way two predictors with different degree of persistence, however which belongs to the same persistence class as defined in KMS.  
In particular, the quantile results we obtain since are based on finite-sample approximations then they have slightly lower value of the true asymptotic quantile that corresponds to the asymptotic distribution of the test. 

\color{black}

\newpage 

\subsection{Power Comparison}

To conduct a power comparison we compare the rejection rates under the alternative hypothesis of structural change, $\mathbb{H}_1: \underline{\beta}_1 \neq \underline{\beta}_2$. Specifically, under the alternative hypothesis, we generate data using the predictive regression given by expression \eqref{eq.break} and then find statistical evidence against the null hypothesis by computing the two test statistics. In particular, comparing the same value of the localizing coefficient across different sample sizes, the power function is indeed monotonically increasing. 

We consider a sequence of local alternatives of the form $\widetilde{\underline{\beta}} = \left( \underline{\beta}^0 + b / T\right)$ and implement the power function for breaks in both the model intercept and the coefficients of the predictors. We introduce parameter instability via the following:
\begin{align*}
\textbf{Design 1} \ \ \ \ \ \ \ \ \ \underline{\beta}^0 &:= \big( \beta_0 = 1, \beta_1 = 1.6, \beta_2 = 0.6  \big)
\\
\\
\textbf{Design 2} \ \ \ \ \ \ \ \ \ \underline{\beta}^0 &:= \big( \beta_0 = 2, \beta_1 = 0.8, \beta_2 =  0.4  \big)
\end{align*}

Table 4 and 5 presents the probabilities of rejection of the two Wald-type statistics at the $5\%$ nominal rate, under the alternative hypothesis. The proposed methodology is recommended in cases in which the practitioner has some prior information regarding the persistent properties of predictors included in the predictive regression. Moreover, for a given value of $c_z$ (e.g., $c_z \equiv 1$), then for a larger exponent rate $( \delta = \delta_0 )$ in the IVX instrument can lead to higher power in exprense of lower values for the empirical size, while a smaller $\delta_0$ can achieve better size corrections. The number of predictors can be another source of distorted inferences and thus in high dimensional settings further corrections might be needed to correct the size and power of the proposed tests. 

The Monte Carlo experiments verify that our proposed IV based Wald test provides a solution to the problem of possible size and power distortions when testing for parameter instability in predictive regressions with persistent predictors. Implementation is therefore straightforward via the use of standard statistical tables. In particular, once the magnitude of the sup-Wald IVX statistic has been computed in the case of a known break-point it then suffices to obtain the relevant critical values from Table 1 in \cite{andrews1993tests} (see e.g., \cite{pitarakis2008comment}).

\begin{remark}
Note that simulating critical values as the results given by Table 1 of \cite{andrews1993tests}, can be done with other numerical approximation methods. Further studies among others include the papers of \cite{estrella2003critical} and \cite{anatolyev2012another}. Moreover, \cite{hansen1997approximate} proposes a methodology to obtain simulated p-values for stability tests in linear regression models, which can be used for the empirical size of the Wald-OLS statistic in our framework. 
\end{remark}



\newpage

\begin{table}[h!]
  \centering
  \caption{Empirical size with nominal level $\alpha = 5\%$.}
    \begin{tabular}{ccrcrccccc}
    \hline
    \multicolumn{10}{c}{\textbf{sup-Wald OLS}} \\
    \hline
    \hline
    \multicolumn{10}{c}{ $\mathbf{c = 1}$  } \\
    \hline
    \textbf{T} & \textbf{-0.9} & \textbf{-0.7} & \textbf{-0.5} & \textbf{-0.3} & \textbf{0} & \textbf{0.3} & \textbf{0.5} & \textbf{0.7} & \textbf{0.9} \\
    \hline
    \textbf{100} & 0.0804 & 0.0736 & 0.0704 & 0.0672 & 0.0556 & 0.0540 & 0.0512 & 0.0664 & 0.0700 \\
    \textbf{250} & 0.0968 & 0.0812 & 0.0700 & 0.0628 & 0.0564 & 0.0608 & 0.0696 & 0.0764 & 0.0904 \\
    \textbf{500} & 0.0924 & 0.0812 & 0.0720 & 0.0644 & 0.0608 & 0.0668 & 0.0744 & 0.0916 & 0.1044 \\
    \textbf{1000} & 0.1060 & 0.0824 & 0.0788 & 0.0716 & 0.0684 & 0.0668 & 0.0760 & 0.0860 & 0.1032 \\
    \hline
    \multicolumn{10}{c}{ $\mathbf{c = 5}$ } \\
    \hline
    \textbf{T} & \textbf{-0.9} & \textbf{-0.7} & \textbf{-0.5} & \textbf{-0.3} & \textbf{0} & \textbf{0.3} & \textbf{0.5} & \textbf{0.7} & \textbf{0.9} \\
    \hline
    \textbf{100} & 0.0588 & 0.0564 & 0.0592 & 0.0548 & 0.0492 & 0.0480 & 0.0500 & 0.0552 & 0.0560 \\
    \textbf{250} & 0.0740 & 0.0608 & 0.0596 & 0.0548 & 0.0516 & 0.0504 & 0.0568 & 0.0692 & 0.0744 \\
    \textbf{500} & 0.0860 & 0.0784 & 0.0696 & 0.0648 & 0.0564 & 0.0572 & 0.0632 & 0.0680 & 0.0848 \\
    \textbf{1000} & 0.0908 & 0.0816 & 0.0732 & 0.0708 & 0.0636 & 0.0596 & 0.0620 & 0.0724 & 0.0864 \\
    \hline
    \multicolumn{10}{c}{ $\mathbf{c = 10}$ } \\
    \hline
    \textbf{T} & \textbf{-0.9} & \textbf{-0.7} & \textbf{-0.5} & \textbf{-0.3} & \textbf{0} & \textbf{0.3} & \textbf{0.5} & \textbf{0.7} & \textbf{0.9} \\
    \hline
    \textbf{100} & 0.0448 & 0.0524 & 0.0508 & 0.0488 & 0.0444 & 0.0448 & 0.0480 & 0.0516 & 0.0492 \\
    \textbf{250} & 0.0572 & 0.0540 & 0.0528 & 0.0488 & 0.0472 & 0.0472 & 0.0548 & 0.0608 & 0.0604 \\
    \textbf{500} & 0.0788 & 0.0768 & 0.0676 & 0.0604 & 0.0544 & 0.0592 & 0.0592 & 0.0648 & 0.0736 \\
    \textbf{1000} & 0.0796 & 0.0768 & 0.0704 & 0.0620 & 0.0568 & 0.0528 & 0.0552 & 0.0652 & 0.0704 \\
    \hline
    \multicolumn{10}{c}{ $\mathbf{c = 20}$ } \\
    \hline
    \textbf{T} & \textbf{-0.9} & \textbf{-0.7} & \textbf{-0.5} & \textbf{-0.3} & \textbf{0} & \textbf{0.3} & \textbf{0.5} & \textbf{0.7} & \textbf{0.9} \\
    \hline
    \textbf{100} & 0.0380 & 0.0416 & 0.0440 & 0.0432 & 0.0476 & 0.0416 & 0.0440 & 0.0440 & 0.0388 \\
    \textbf{250} & 0.0560 & 0.0476 & 0.0480 & 0.0408 & 0.0436 & 0.0460 & 0.0496 & 0.0480 & 0.0484 \\
    \textbf{500} & 0.0620 & 0.0592 & 0.0600 & 0.0560 & 0.0512 & 0.0504 & 0.0564 & 0.0596 & 0.0648 \\
    \textbf{1000} & 0.0676 & 0.0656 & 0.0608 & 0.0548 & 0.0496 & 0.0520 & 0.0548 & 0.0576 & 0.0572 \\
    \hline
    \hline
    \end{tabular}%
  \label{Table1}%
\end{table}%

\begin{small}
Table \ref{Table1} presents finite-sample empirical sizes for the sup Wald-OLS test, with nominal level $\alpha = 5\%$ for $B = 5,000$ replications. The predictive regression model under the null hypothesis, $H_0: \beta_1 = \beta_2$, is given by, $y_{t} = 0.25 x_{t-1} + u_t$, $x_t = (1- \frac{c}{T}) x_{t-1} + v_t$, with $\Sigma_{ee} = \begin{bmatrix}
1 & \rho \\
\rho    & 1
\end{bmatrix}$. 
\end{small}

\newpage 

\begin{table}[h!]
  \centering
  \caption{Empirical size with nominal level $\alpha = 5\%$.}
   \begin{tabular}{ccrcrccccc}
    \hline
    \multicolumn{10}{c}{\textbf{sup-Wald OLS}} \\
    \hline
    \hline
    \multicolumn{10}{c}{ $\mathbf{c = -1}$ } \\
    \hline
    \textbf{T} & \textbf{-0.9} & \textbf{-0.7} & \textbf{-0.5} & \textbf{-0.3} & \textbf{0} & \textbf{0.3} & \textbf{0.5} & \textbf{0.7} & \textbf{0.9} \\
    \hline
    \textbf{100} & 0.0692 & 0.0636 & 0.0656 & 0.0600 & 0.0584 & 0.0604 & 0.0584 & 0.0616 & 0.0616 \\
    \textbf{250} & 0.0724 & 0.0616 & 0.0596 & 0.0660 & 0.0584 & 0.0636 & 0.0640 & 0.0668 & 0.0748 \\
    \textbf{500} & 0.0804 & 0.0680 & 0.0664 & 0.0604 & 0.0612 & 0.0628 & 0.0720 & 0.0740 & 0.0868 \\
    \textbf{1000} & 0.0728 & 0.0636 & 0.0700 & 0.0680 & 0.0636 & 0.0644 & 0.0680 & 0.0764 & 0.0768 \\
    \hline
    \multicolumn{10}{c}{ $\mathbf{c = -5}$ } \\
    \hline
    \textbf{T} & \textbf{-0.9} & \textbf{-0.7} & \textbf{-0.5} & \textbf{-0.3} & \textbf{0} & \textbf{0.3} & \textbf{0.5} & \textbf{0.7} & \textbf{0.9} \\
    \hline
    \textbf{100} & 0.0660 & 0.0672 & 0.0732 & 0.0740 & 0.0664 & 0.0732 & 0.0732 & 0.0696 & 0.0656 \\
    \textbf{250} & 0.0700 & 0.0756 & 0.0804 & 0.0832 & 0.0836 & 0.0820 & 0.0812 & 0.0820 & 0.0816 \\
    \textbf{500} & 0.0692 & 0.0660 & 0.0752 & 0.0784 & 0.0792 & 0.0856 & 0.0836 & 0.0772 & 0.0764 \\
    \textbf{1000} & 0.0804 & 0.0880 & 0.0916 & 0.0896 & 0.0864 & 0.0828 & 0.0864 & 0.0748 & 0.0708 \\
    \hline
    \multicolumn{10}{c}{ $\mathbf{c = -10}$ } \\
    \hline
    \textbf{T} & \textbf{-0.9} & \textbf{-0.7} & \textbf{-0.5} & \textbf{-0.3} & \textbf{0} & \textbf{0.3} & \textbf{0.5} & \textbf{0.7} & \textbf{0.9} \\
    \hline
    \textbf{100} & 0.1124 & 0.1084 & 0.1060 & 0.1072 & 0.1020 & 0.1056 & 0.1104 & 0.1176 & 0.1160 \\
    \textbf{250} & 0.1264 & 0.1396 & 0.1384 & 0.1372 & 0.1244 & 0.1244 & 0.1276 & 0.1304 & 0.1368 \\
    \textbf{500} & 0.1220 & 0.1224 & 0.1260 & 0.1260 & 0.1240 & 0.1280 & 0.1288 & 0.1376 & 0.1432 \\
    \textbf{1000} & 0.1528 & 0.1468 & 0.1424 & 0.1340 & 0.1308 & 0.1272 & 0.1316 & 0.1384 & 0.1400 \\
    \hline
    \multicolumn{10}{c}{ $\mathbf{c = -20}$ } \\
    \hline
    \textbf{T} & \textbf{-0.9} & \textbf{-0.7} & \textbf{-0.5} & \textbf{-0.3} & \textbf{0} & \textbf{0.3} & \textbf{0.5} & \textbf{0.7} & \textbf{0.9} \\
    \hline
    \textbf{100} & 0.1576 & 0.1500 & 0.1464 & 0.1532 & 0.1564 & 0.1532 & 0.1584 & 0.1632 & 0.1704 \\
    \textbf{250} & 0.2104 & 0.2036 & 0.1988 & 0.2004 & 0.1908 & 0.1976 & 0.1956 & 0.2016 & 0.1928 \\
    \textbf{500} & 0.2144 & 0.2144 & 0.2040 & 0.2024 & 0.2040 & 0.2176 & 0.2200 & 0.2224 & 0.2460 \\
    \textbf{1000} & 0.2520 & 0.2552 & 0.2368 & 0.2376 & 0.2284 & 0.2320 & 0.2336 & 0.2360 & 0.2528 \\
    \hline
    \hline
    \end{tabular}%
  \label{Table1A}%
\end{table}%

\begin{small}
Table \ref{Table1A} presents finite-sample empirical sizes for the sup Wald-OLS test, with nominal level $\alpha = 5\%$ for $B = 5,000$ replications. The predictive regression model under the null hypothesis, $H_0: \beta_1 = \beta_2$, is given by, $y_{t} = 0.25 x_{t-1} + u_t$, $x_t = (1- \frac{c}{T}) x_{t-1} + v_t$, with $\Sigma_{ee} = \begin{bmatrix}
1    & \rho \\
\rho & 1
\end{bmatrix}$. 
\end{small}


\newpage 

\color{red}
\textbf{OLD TABLE} (to update with the ones after bootstrapping)
\color{black}

\begin{table}[h!]
  \centering
  \caption{Empirical size with nominal level $\alpha = 5\%$.}
    \begin{tabular}{rrrrrrrrrr}
    \hline
    \multicolumn{10}{c}{\textbf{sup Wald-IVX statistic}} \\
     \hline
      \hline
    \multicolumn{10}{c}{ \textbf{c = 1} ($c_z = 1$, $\delta = 0.75$, $c_{\alpha} = 13.42$) } \\
     \hline
    \multicolumn{1}{c}{\textbf{T}} & \multicolumn{1}{c}{\textbf{-0.9}} & \multicolumn{1}{c}{\textbf{-0.7}} & \multicolumn{1}{c}{\textbf{-0.5}} & \multicolumn{1}{c}{\textbf{-0.3}} & \multicolumn{1}{c}{\textbf{0}} & \multicolumn{1}{c}{\textbf{0.3}} & \multicolumn{1}{c}{\textbf{0.5}} & \multicolumn{1}{c}{\textbf{0.7}} & \multicolumn{1}{c}{\textbf{0.9}} \\
     \hline
    \multicolumn{1}{c}{\textbf{100}} & \multicolumn{1}{c}{0.1672} & \multicolumn{1}{c}{0.1070} & \multicolumn{1}{c}{0.0718} & \multicolumn{1}{c}{0.0480} & \multicolumn{1}{c}{0.0368} & \multicolumn{1}{c}{0.0496} & \multicolumn{1}{c}{0.0710} & \multicolumn{1}{c}{0.1106} & \multicolumn{1}{c}{0.1726} \\
    \multicolumn{1}{c}{\textbf{250}} & \multicolumn{1}{c}{0.1336} & \multicolumn{1}{c}{0.0878} & \multicolumn{1}{c}{0.0486} & \multicolumn{1}{c}{0.0324} & \multicolumn{1}{c}{0.0232} & \multicolumn{1}{c}{0.0322} & \multicolumn{1}{c}{0.0506} & \multicolumn{1}{c}{0.0838} & \multicolumn{1}{c}{0.1354} \\
    \multicolumn{1}{c}{\textbf{500}} & \multicolumn{1}{c}{0.1316} & \multicolumn{1}{c}{0.0808} & \multicolumn{1}{c}{0.0504} & \multicolumn{1}{c}{0.0306} & \multicolumn{1}{c}{0.0206} & \multicolumn{1}{c}{0.0282} & \multicolumn{1}{c}{0.0462} & \multicolumn{1}{c}{0.0772} & \multicolumn{1}{c}{0.1226} \\
    \multicolumn{1}{c}{\textbf{1000}} & \multicolumn{1}{c}{0.1076} & \multicolumn{1}{c}{0.0644} & \multicolumn{1}{c}{0.0412} & \multicolumn{1}{c}{0.0264} & \multicolumn{1}{c}{0.0200} & \multicolumn{1}{c}{0.0260} & \multicolumn{1}{c}{0.0398} & \multicolumn{1}{c}{0.0668} & \multicolumn{1}{c}{0.1044} \\
     \hline
    \multicolumn{10}{c}{ \textbf{c = 5} ($c_z = 1$, $\delta = 0.75$, $c_{\alpha} = 13.42$) } \\
     \hline
    \multicolumn{1}{c}{\textbf{T}} & \multicolumn{1}{c}{\textbf{-0.9}} & \multicolumn{1}{c}{\textbf{-0.7}} & \multicolumn{1}{c}{\textbf{-0.5}} & \multicolumn{1}{c}{\textbf{-0.3}} & \multicolumn{1}{c}{\textbf{0}} & \multicolumn{1}{c}{\textbf{0.3}} & \multicolumn{1}{c}{\textbf{0.5}} & \multicolumn{1}{c}{\textbf{0.7}} & \multicolumn{1}{c}{\textbf{0.9}} \\
     \hline
    \multicolumn{1}{c}{\textbf{100}} & \multicolumn{1}{c}{0.1134} & \multicolumn{1}{c}{0.0804} & \multicolumn{1}{c}{0.0580} & \multicolumn{1}{c}{0.0426} & \multicolumn{1}{c}{0.0350} & \multicolumn{1}{c}{0.0446} & \multicolumn{1}{c}{0.0596} & \multicolumn{1}{c}{0.0850} & \multicolumn{1}{c}{0.1210} \\
    \multicolumn{1}{c}{\textbf{250}} & \multicolumn{1}{c}{0.0998} & \multicolumn{1}{c}{0.0662} & \multicolumn{1}{c}{0.0456} & \multicolumn{1}{c}{0.0348} & \multicolumn{1}{c}{0.0248} & \multicolumn{1}{c}{0.0326} & \multicolumn{1}{c}{0.0452} & \multicolumn{1}{c}{0.0666} & \multicolumn{1}{c}{0.0986} \\
    \multicolumn{1}{c}{\textbf{500}} & \multicolumn{1}{c}{0.0944} & \multicolumn{1}{c}{0.0592} & \multicolumn{1}{c}{0.0410} & \multicolumn{1}{c}{0.0250} & \multicolumn{1}{c}{0.0210} & \multicolumn{1}{c}{0.0282} & \multicolumn{1}{c}{0.0448} & \multicolumn{1}{c}{0.0628} & \multicolumn{1}{c}{0.0916} \\
    \multicolumn{1}{c}{\textbf{1000}} & \multicolumn{1}{c}{0.0752} & \multicolumn{1}{c}{0.0488} & \multicolumn{1}{c}{0.0318} & \multicolumn{1}{c}{0.0236} & \multicolumn{1}{c}{0.0190} & \multicolumn{1}{c}{0.0250} & \multicolumn{1}{c}{0.0364} & \multicolumn{1}{c}{0.0522} & \multicolumn{1}{c}{0.0768} \\
     \hline
      \hline
          &       &       &       &       &       &       &       &       &  \\
           \hline
            \hline
       \multicolumn{10}{c}{ \textbf{c = 1} ($c_z = 1$, $\delta = 0.75$, $c_{\alpha} = 12$)     } \\
        \hline
    \multicolumn{1}{c}{\textbf{T}} & \multicolumn{1}{c}{\textbf{-0.9}} & \multicolumn{1}{c}{\textbf{-0.7}} & \multicolumn{1}{c}{\textbf{-0.5}} & \multicolumn{1}{c}{\textbf{-0.3}} & \multicolumn{1}{c}{\textbf{0}} & \multicolumn{1}{c}{\textbf{0.3}} & \multicolumn{1}{c}{\textbf{0.5}} & \multicolumn{1}{c}{\textbf{0.7}} & \multicolumn{1}{c}{\textbf{0.9}} \\
     \hline
    \multicolumn{1}{c}{\textbf{100}} & \multicolumn{1}{c}{0.2386} & \multicolumn{1}{c}{0.1610} & \multicolumn{1}{c}{0.1108} & \multicolumn{1}{c}{0.0768} & \multicolumn{1}{c}{0.0632} & \multicolumn{1}{c}{0.0814} & \multicolumn{1}{c}{0.1138} & \multicolumn{1}{c}{0.1612} & \multicolumn{1}{c}{0.2482} \\
    \multicolumn{1}{c}{\textbf{250}} & \multicolumn{1}{c}{0.2004} & \multicolumn{1}{c}{0.1336} & \multicolumn{1}{c}{0.0816} & \multicolumn{1}{c}{0.0538} & \multicolumn{1}{c}{0.0390} & \multicolumn{1}{c}{0.0558} & \multicolumn{1}{c}{0.0822} & \multicolumn{1}{c}{0.1338} & \multicolumn{1}{c}{0.2042} \\
    \multicolumn{1}{c}{\textbf{500}} & \multicolumn{1}{c}{0.1938} & \multicolumn{1}{c}{0.1280} & \multicolumn{1}{c}{0.0828} & \multicolumn{1}{c}{0.0538} & \multicolumn{1}{c}{0.0360} & \multicolumn{1}{c}{0.0536} & \multicolumn{1}{c}{0.0792} & \multicolumn{1}{c}{0.1232} & \multicolumn{1}{c}{0.1900} \\
    \multicolumn{1}{c}{\textbf{1000}} & \multicolumn{1}{c}{0.1644} & \multicolumn{1}{c}{0.1048} & \multicolumn{1}{c}{0.0672} & \multicolumn{1}{c}{0.0448} & \multicolumn{1}{c}{0.0356} & \multicolumn{1}{c}{0.0454} & \multicolumn{1}{c}{0.0686} & \multicolumn{1}{c}{0.1048} & \multicolumn{1}{c}{0.1664} \\
     \hline
    \multicolumn{10}{c}{\textbf{c = 5} ($c_z = 1$, $\delta = 0.75$, $c_{\alpha} = 12$) }  \\
     \hline
    \multicolumn{1}{c}{\textbf{T}} & \multicolumn{1}{c}{\textbf{-0.9}} & \multicolumn{1}{c}{\textbf{-0.7}} & \multicolumn{1}{c}{\textbf{-0.5}} & \multicolumn{1}{c}{\textbf{-0.3}} & \multicolumn{1}{c}{\textbf{0}} & \multicolumn{1}{c}{\textbf{0.3}} & \multicolumn{1}{c}{\textbf{0.5}} & \multicolumn{1}{c}{\textbf{0.7}} & \multicolumn{1}{c}{\textbf{0.9}} \\
     \hline
    \multicolumn{1}{c}{\textbf{100}} & \multicolumn{1}{c}{0.1722} & \multicolumn{1}{c}{0.1234} & \multicolumn{1}{c}{0.0936} & \multicolumn{1}{c}{0.0704} & \multicolumn{1}{c}{0.0568} & \multicolumn{1}{c}{0.0752} & \multicolumn{1}{c}{0.0966} & \multicolumn{1}{c}{0.1332} & \multicolumn{1}{c}{0.1818} \\
    \multicolumn{1}{c}{\textbf{250}} & \multicolumn{1}{c}{0.1588} & \multicolumn{1}{c}{0.1094} & \multicolumn{1}{c}{0.0714} & \multicolumn{1}{c}{0.0548} & \multicolumn{1}{c}{0.0458} & \multicolumn{1}{c}{0.0544} & \multicolumn{1}{c}{0.0766} & \multicolumn{1}{c}{0.1080} & \multicolumn{1}{c}{0.1548} \\
    \multicolumn{1}{c}{\textbf{500}} & \multicolumn{1}{c}{0.1476} & \multicolumn{1}{c}{0.0984} & \multicolumn{1}{c}{0.0716} & \multicolumn{1}{c}{0.0506} & \multicolumn{1}{c}{0.0366} & \multicolumn{1}{c}{0.0494} & \multicolumn{1}{c}{0.0708} & \multicolumn{1}{c}{0.1008} & \multicolumn{1}{c}{0.1422} \\
    \multicolumn{1}{c}{\textbf{1000}} & \multicolumn{1}{c}{0.1190} & \multicolumn{1}{c}{0.0828} & \multicolumn{1}{c}{0.0578} & \multicolumn{1}{c}{0.0428} & \multicolumn{1}{c}{0.0358} & \multicolumn{1}{c}{0.0456} & \multicolumn{1}{c}{0.0630} & \multicolumn{1}{c}{0.0898} & \multicolumn{1}{c}{0.1276} \\
     \hline
     \hline
    \end{tabular}%
  \label{Table2}%
\end{table}%

\begin{small}
Table \ref{Table2} presents finite-sample sizes for the sup Wald-IVX test, with nominal size $\alpha = 5\%$. The predictive regression model under the null hypothesis is given by $y_{t} = 0.25 + 0.5 x_{t-1} + u_t$, $x_t = (1- \frac{c_1}{T}) x_{t-1} + v_t$, with $\Sigma_{ee} = \begin{bmatrix}
0.25 & \sigma_{uv} \\
\sigma_{uv}    & 0.75
\end{bmatrix}$, $\rho = \displaystyle \frac{ \sigma_{uv}  }{ \sigma_u \sigma_v }$, as given above.  Furthermore, for the IVX estimation step, we use an IVX persistence parameter $\delta \in \left\{ 0.75, 0.95 \right\}$ and the localizing coefficient is set to $c_{z} = 1$. The number of replications is $B = 5,000$.
\end{small}

\newpage 






\section{Empirical Application}
\label{Section5}

In this section we present an empirical application aiming to shed light on the literature of stock return predictability.  Identifying periods of predictability has important implications in various aspects of finance. Related modern reviews of these aspects are presented by \cite{kostakis2018taking} and \cite{chinco2019sparse}, among others. However, despite the extensive research of the field, the findings are still rather mixed \citep{kasparis2015nonparametric} (see, \cite{welch2008comprehensive} for a full discussion). For instance, aspects such as the chosen sample period or the selected predictors can give different conclusions. Additionally, parameter instability due to certain economic events can also affect the reliability of predictability tests. Therefore, it is of paramount importance to develop robust testing methodologies for inferring predictability under conditions such as parameter instability or the presence of nonstationary regressors. Using the predictive regression model studied in this paper\footnote{Alternative model specifications can be considered; for instance a model which considers expected returns in relation to macroeconomic conditions and forecasting uncertainty. A first move towards this direction is presented by \cite{atanasov2020consumption} who examine consumption fluctuations and expected returns with respect to the predictability literature. More specifically, the authors indeed find statistical evidence of predictability at the one-quarter horizon using the IVX testing approach of KMS.} our primary focus is to examine the robustness of the proposed tests. 

\subsection{Data Description} 

We focus on examining the presence of predictability for monthly US stock market excess returns over the period 1990-2019 using the set of variables considered in \cite{welch2008comprehensive}\footnote{The dataset can be retrieved from Amit Goyal's website at \url{http://www.hec.unil.ch/agoyal/}. Detailed descriptions of variables can be found in the Online Appendix of \cite{welch2008comprehensive}. } which capture economic and financial conditions for the US economy.  

\paragraph{Predictant}  The dependent variable is the monthly equity premium (excess return) of the US stock market based on the S$\&$P500  index. We construct the excess return as in \cite{kasparis2015nonparametric}, that is, the difference between the total rate of return and the risk-free rate for the same sample period. As a proxy of the US stock market return, we use the value-weighted S$\&$P500 total stock market return including dividends. The risk-free rate is the 3-month T-bill rate obtained from the database of FRED\footnote{Time series of macroeconomic variables, such as the US inflation rate and the T-bill rate can be found at \url{ https://fred.stlouisfed.org/}. Notice also that the proxy of equity premium and the other financial variables we consider in this paper, are commonly used in the predictability literature, see  \cite{gonzalo2012regime}, \cite{kasparis2015nonparametric},\cite{kostakis2015Robust} and \cite{kostakis2018taking}.}.

\newpage 

\paragraph{Predictors} The predictor variables we consider include: 
dividend-payout ratio (d/e), long-term yield (lty), dividend yield spread (dy), dividend-price-ratio (d/p), T-bill rate (tbl), earnings-price-ratio (e/p), book-to-market ratio (b/m), default yield spread (dfs), net equity expansion (ntis), term spread (tms) and inflation rate (inf). 

\subsection{Predictability Tests}

To begin with, we apply the simple predictability test on the full sample for the period 1946-2019 using as predictant the S$\&$P500 Equity Premium.  However, in this paper we take a slightly different approach than the literature. In particular, we focus on the subsample spanning the period 1990Q1 to 2019Q4, and consider monthly sampling frequency. Our first goal is to examine the stock return predictability of this subsample, that is, to identify the financial variables which are individually statistical significant as well as to identify for evidence of joint statistical significance. Furthermore, our second goal is to repeat the same exercise using the proposed joint test of predictability and parameter instability and compare the results we obtain. The chosen subsample includes the period of the 2008 financial crisis, so it is natural to assume that certain predictors might exhibit structural break around that economic event. Therefore, this is a suitable sample to assess the statistical validity of the proposed methodology for the case of a single structural break. Certain limitations of our approach are on sight, however these do not invalidate our findings. In particular, we do not consider the existence of multiple structural breaks neither we consider sample splitting techniques which can affect the power of the tests especially when using an out-of-sample forecasting scheme.  

Firstly, summarizing our findings is Table \ref{tableB1} which presents predictability tests based on both the classical least squares estimator and the IVX estimator. Notice that for these set of tests we consider the regressors on their original form in order to preserve the degree of persistence and have comparability between the two estimators under examination. Table  \ref{tableB2} presents structural break tests for the regressors. In this case the traditional structural break tests are implemented under the assumption of stationary time series, by taking the first difference of the regressors before fitting the AR(1) models.  

Secondly, we examine the short-horizon predictability via the proposed joint predictability and structural break Wald statistics.

\newpage 

\section{Conclusion}
\label{Section6}

In this paper, we have extensively examined the asymptotic theory of tests for Joint predictability and parameter instability under the assumption of nonstationary regressors. We compare our results with previous seminal work in the fields of both structural break testing in linear regressions as well as predictability testing in predictive regression models. We find some interesting results not previously presented in the literature. Firstly, using the OLS estimator for the parameters of the predictive regression model we show that the limiting distribution of the Wald statistic for testing for a single structural break has a nonstandard limiting distribution which depends on the unknown coefficient of persistence. Secondly, by employing the IVX estimator proposed in the literature as a robust estimator which filters out the abstract degree of persistence in regressors, we have proved that the limiting distribution of the Joint tests takes different forms which weakly converges to a functional of a Brownian bridge in some instances while converges to a nonstandard limiting distribution which depends on the coefficient of persistence when regressors are assume to be highly persistent.


Conducting inference, such as structural break testing, on the regression coefficient of the predictive regression model with multiple highly persistent regressors can lead to a nonstandard limiting distribution. The proposed testing methodology ensures that the limiting distribution of the structural break tests is free of any nuisance parameters, such as the unknown localizing coefficient of persistence.  Moreover, we consider "pure" structural change as it is defined by \cite{andrews1993tests}, in the sense that the entire parameter vector is subject to structural change under the alternative hypothesis. The asymptotic distribution of the IV based Wald test is found to be given by the supremum of the normalized squared Brownian Bridge (NBB). Thus, the exact limiting distribution of the IV based Wald test allows to determine critical values similar to the case of the linear regression model, without further simulations and additional computational cost. This holds in the case of mildly integrated or integrated regressors, while in the case of nonstationary regressors further investigation is needed to determine exact critical values since the limiting distribution has a dependence on the nuisance coefficient of persistence.  

The developed asymptotic theory for the sup-Wald IVX tests  under various degrees of persistence we consider in this paper, indicate that the asymptotic behaviour of the tests when the supremum functional is included, is different from the corresponding limit theory in the case of only linear restrictions to the parameters of the predictive regression. Nevertheless, the robustness of the IVX instrumentation provides a way to determine an analytic form of the asymptotic distribution for different levels of persistence, a feature often seen in time series data. This feature appears in various empirical finance applications in which the available information set for current or future economic conditions with many times persistent properties and existence of parameter instability.

\newpage

\newpage 

\bibliographystyle{apalike}
\bibliography{myreferences1}

\newpage 

\appendix

\newpage 

\section{Asymptotic Theory}
\label{AppendixA}

\renewcommand{\thetable}{\Alph{section}\arabic{table}}
\setcounter{table}{0}

In this Appendix we present the main mathematical derivations and proofs related to the results reported in the main body. We begin by summarizing via Lemma \ref{lemma1} below the limit theory results which can be found in \cite{phillipsmagdal2009econometric} and \cite{kostakis2015Robust}. We introduce the shorthand notation $\alpha \wedge \beta \equiv \text{min}( \alpha, \beta )$ to denote the minimum operator, employed for the stochastic dominance of the convergence rates.


\medskip

\begin{lemma}
\label{lemma1}
Let $\mathbb{V}_{xz} := \displaystyle \int_0^{\infty} e^{rC} \mathbf{V}_{xx} e^{rC_z} dr$, where $\mathbf{V}_{xx} := \displaystyle \int_0^{\infty} e^{s C} \mathbf{\Omega}_{xx} e^{s C} ds$, and $\mathbf{\Omega}_{xx}$ is the long-run covariance of $u_{t}$. Then, under the null hypothesis of no structural break in the predictive regression model, the following asymptotic results hold: 
\begin{enumerate}
\item[\textit{(i)}] the sample covariance satisfies that 
\begin{align}
\frac{1}{T^{ \frac{ 1 + \gamma_x \wedge \delta_z }{2} } }
 \sum_{t=1}^{ \floor{T \pi } } \widetilde{z}_{t-1} \big( u_{t} - \bar{u}_T \big) \Rightarrow U \left( \pi \right)
\end{align}
where $U \left( . \right)$ is a Brownian motion with variance $\sigma^2_u \widetilde{\mathbf{V}}$, where $\widetilde{\mathbf{V}}$ is defined as
\begin{equation}
\widetilde{\mathbf{V}} =
\begin{cases}
\displaystyle \int_0^{\infty} e^{r C_z} \mathbf{\Omega}_{xx} e^{r C_z} dr 
& ,\text{if} \ \gamma_x > \delta_z 
\\
\\
\displaystyle \int_0^{\infty} e^{r C_z} \left( \mathbf{C} \mathbb{V}_{xz} + \mathbf{C}_z \mathbb{V}_{xz}^{\prime} \mathbf{C} \right) e^{r C_z} dr   
& ,\text{if} \ \gamma_x = \delta_z 
\\
\\
\displaystyle \int_0^{\infty} e^{r C} \mathbf{\Omega}_{xx} e^{r C} dr   
& ,\text{if} \ 0 < \gamma_x < \delta_z  
\\
\\
\mathbb{E} \left( x_{0,1} x_{0,1}^{\prime} \right)
& ,\text{if} \ \gamma_x = 0.  
\end{cases}
\end{equation}
\end{enumerate} 
where $x_{0,t} = \sum_{ j=0 }^{ \infty } \left( \mathbf{I}_p + \mathbf{C} \right)^j u_{t-j}$ is the corresponding stationary sequence of the regressor vector $x_t$ when $\gamma_x = 0$.
\item[\textit{(ii)}] the sample second moment satisfies that 
\begin{align}
\frac{1}{ T^{ \big( 1 + \gamma_x \wedge \delta_z \big)}} 
 \sum_{t=1}^{ \floor{ T \pi} } \widetilde{z}_{t-1} \big( x_{t-1} - \bar{x}_{T-1} \big)^{\prime} \Rightarrow \mathbf{\Psi} \left( \pi \right)
\end{align}

\medskip

where $\mathbf{\Psi} \left( \pi \right)$ has a different asymptotic convergence result as below, depending on the exponent rates $\gamma_x$ and $\delta_z$ of the original regressor and instrumental regressor respectively. 

\newpage 

\begin{equation}
\mathbf{\Psi} \left( \pi \right) =
\begin{cases}
\displaystyle - \mathbf{C}_z^{-1} \left( \pi \mathbf{\Omega}_{xx} +  \int_0^{\pi} \underline{ \mathbf{B} } dB^{\prime} \right)    & ,\text{if} \ \gamma_x > 1 
\\
\\
\displaystyle - \mathbf{C}_z^{-1} \left( \pi \mathbf{\Omega}_{xx} +  \int_0^{ \pi } \underline{ \mathbf{J} }_C dJ_C^{\prime} \right)    & ,\text{if} \ \gamma_x = 1 
\\
\\
\displaystyle - \pi \mathbf{C}_z^{-1} \bigg( \mathbf{\Omega}_{xx} + \int_0^{\infty} e^{rC} \mathbf{\Omega}_{xx} e^{rC} dr \mathbf{C}  \bigg)   
& ,\text{if} \ \delta_x < \gamma_x < 1  
\\
\\
\displaystyle - \pi \mathbf{C} \mathbb{V}_{xz}   
& ,\text{if} \ \gamma_x = \delta_x  
\\
\\
\displaystyle \pi \int_0^{ \infty } e^{rC} \mathbf{\Omega}_{xx} e^{rC} dr   
& ,\text{if} \ 0 < \gamma_x < \delta_x    
\\
\\
\displaystyle \pi \mathbb{E} \left( x_{0,1} x_{0,1}^{\prime} \right)   
& ,\text{if} \ \gamma_x = 0. 
\end{cases}
\end{equation}
where $B(.)$ is a $p-$dimensional standard Brownian motion, $J_C (\pi) = \int_0^{\pi} e^{C (\pi - s)} dB(\pi)$ is an \textit{Ornstein-Uhkenbeck} (OU) process and we denote with $\underline{J}_C (\pi) = J_C (\pi) - \int_0^1 J_C(s) ds$ and $\underline{B} ( \pi ) = B(\pi) - \int_0^1 B(s) ds$ the demeaned processes of $J(\pi)$ and $B(\pi)$ respectively.
\item[\textit{(iii)}] The weakly joint convergence result applies and the asymptotic terms given by expressions in \textit{(i)} and \textit{(ii)} are stochastically independent.  
\end{lemma}

Notice that for summarizing the above results we used that 
\begin{align}
\frac{1}{ T^{1 + \delta_z } }  \sum_{t=1}^T z_{t-1} z_{t-1}^{\prime}  \overset{ \text{plim} }{ \to } \mathbf{V}_{zz} :=  \int_0^{\infty} e^{rC_z} \mathbf{\Omega}_{xx} e^{rC_z} dr  
\end{align}
Moreover, we have the weakly convergence result from \cite{phillipsmagdal2009econometric}:
\begin{align}
\label{mixed.gaussian}
\frac{1}{ T^{ \frac{ 1 + \delta_z }{ 2 } } } \sum_{t=1}^T \left( z_{t-1} \otimes u_t  \right)  
\Rightarrow \mathcal{N} \big( 0 , \mathbf{V}_{zz} \otimes \mathbf{\Sigma}_{ uu } \big)   
\end{align}
Expression \eqref{mixed.gaussian} proves a mixed Gaussian limiting distribution. This, shows that the limit distribution of $T^{ - (1 + \delta_z )/2} \sum_{t=1}^T \left( z_{t-1} \otimes u_t  \right)$ is Gaussian with mean zero and covariance matrix equal to the probability limit of $T^{ - (1 + \delta )/2} \sum_{t=1}^T \left( z_{t-1} \otimes u_t  \right)$, which is equal to $\mathbf{V}_{zz} \otimes \mathbf{\Sigma}_{ uu }$, where $\mathbf{V}_{zz} := \int_0^{\infty} e^{r C_z} \mathbf{\Omega}_{xx} e^{r C_z} dr$. Specifically, the above Mixed Gaussianity convergence, is a powerful result within the IVX framework and ensures the robustness of the methodology and the estimation procedure. The dependence of the covariance matrix on the degree of persistence of the IVX instrumentation methodology, induces exactly the Mixed Gaussianity. Similarly, the limit distribution of $T^{ - (1 + \delta_z )/2} \sum_{t=1}^T \big( x_{t-1} \otimes u_t  \big)  
\Rightarrow \mathcal{N} \big( 0 , \mathbf{V}_{xx} \otimes \mathbf{\Sigma}_{ uu } \big)$, where $\mathbf{V}_{xx} := \int_0^{\infty} e^{r C} \mathbf{\Omega}_{xx} e^{r C} dr$, is proved in Lemma 3.3 of PM.

\newpage 

\textbf{Proof of Theorem \ref{Theorem1}. }  

\begin{proof}

We denote with  $\widetilde{x}_t = \left(1, x_t^{\prime} \right)^{\prime}$  and with $\theta_j = (\alpha_j, \beta_j)^{\prime}$ for $j=1,2$ the parameter vector which is obtained via the OLS estimator. Then, under the null hypothesis of no structural break, $\mathbb{H}_0: \theta_1 = \theta_2$, against $\mathbb{H}_1: \theta_1 \neq \theta_2$, we obtain the following expressions 
\begin{align*}
\left(  \widehat{\theta}_{1} - \theta^0 \right) 
&= 
\left(  \sum_{t=1}^T \widetilde{ x }_{t-1} \widetilde{ x }_{t-1}^{\prime} I_{1t} \right)^{-1}  \left( \sum_{t=1}^T \widetilde{ x }_{t-1} u_{t}  I_{1t} \right) 
\\
\left(  \widehat{\theta}_{2} - \theta^0 \right) 
&= 
\left(  \sum_{t=1}^T \widetilde{ x }_{t-1} \widetilde{ x }_{t-1}^{\prime} I_{2t} \right)^{-1}  \left( \sum_{t=1}^T \widetilde{ x }_{t-1} u_{t}  I_{2t} \right) 
\end{align*}
with $I_{1t}$ and $I_{2t}$ the dummy time variables and  $\theta^0 = ( \alpha_0, \beta_0 )^{\prime}$, the population value of the parameter vector $\theta$.  Therefore, we have that 
\begin{align*}
T \left( \widehat{\theta}_{1} - \theta^0 \right)  
&= \left( \frac{1}{T^2} \sum_{t=1}^{ \floor{ T \pi} } \widetilde{x}_{t-1} \widetilde{x}_{t-1}^{\prime} \right)^{-1} \left( \frac{1}{T} \sum_{t=1}^{ \floor{ T \pi} } \widetilde{x}^{\prime}_{t-1} u_{t} \right) 
\\
T \left( \widehat{\theta}_{2} - \theta^0 \right)  
&= \left( \frac{1}{T^2} \sum_{t= \floor{ T \pi} + 1 }^{ T } \widetilde{x}_{t-1} \widetilde{x}_{t-1}^{\prime} \right)^{-1} \left( \frac{1}{T} \sum_{t=1}^{ \floor{ T \pi} } \widetilde{x}^{\prime}_{t-1} u_{t} \right) 
\end{align*}
Then, the weakly convergence result for the estimator of $\beta_1$ follows
\begin{align}
\label{expression6}
T \left( \widehat{\theta}_{1} - \theta^0 \right) 
&\Rightarrow \left( \int_0^{\pi} \widetilde{K}_c(r) \widetilde{K}^{\prime}_c(r) dr \right)^{-1}  \left( \int_0^{\pi} \widetilde{K}_c(r) d B_u \right) 
\end{align}
Similarly, for the estimator of $\beta_2$ we have the following weakly convergence result
\begin{align}
\label{expression7}
T \left( \widehat{\theta}_{2} - \theta^0 \right) 
&\Rightarrow \left( \int_{\pi}^1 \widetilde{K}_c(r) \widetilde{K}^{\prime}_c(r) dr \right)^{-1}  \left( \int_{\pi}^1 \widetilde{K}_c(r) d B_u \right) 
\end{align}
In order to simplify the expression of the Wald OLS statistic we denote with 
\begin{align}
\widetilde{G}_c(\pi) := \int_0^{\pi} \widetilde{ K }_c( r ) \widetilde{ K}^{\prime}_c( r ) dr \ \ \text{and} \ \  H_c( \pi ) := \int_0^{\pi} \widetilde{ K }_c( r ) d B_u(r) 
\end{align}
which implies that due to the argument $\pi$ in the expressions for $ \widetilde{G}_c(\pi)$ and $\widetilde{ H }_c( \pi )$ 
\begin{align}
\widetilde{G}_c(1) := \int_0^{1} \widetilde{ K }_c( r ) \widetilde{ K }^{\prime}_c( r ) dr \ \ \text{and} \ \  \widetilde{ H }_c( 1 ) := \int_0^{1} \widetilde{ K }_c( r ) d B_u(r) 
\end{align}
Notice that, for example we can deduce that 
\begin{align*}
\left( \int_{\pi}^1 \widetilde{K}_c(r) \widetilde{K}^{\prime}_c(r) dr \right) 
&= \left( \int_{0}^1 \widetilde{K}_c(r) \widetilde{K}^{\prime}_c(r) dr \right) - \left( \int_0^{\pi} \widetilde{K}_c(r) \widetilde{K}^{\prime}_c(r) dr \right) :=  \widetilde{\textbf{G}}_c(1) - \widetilde{\textbf{G}}_c( \pi )
\end{align*}

\newpage 

Thus, the statistical distance component of the sup Wald-OLS statistic is given by  
\begin{align}
T  \left( \widehat{\theta}_{1} - \widehat{\theta}_{2} \right) 
= 
\left\{ \widetilde{\textbf{G}}_c(\pi)^{-1}  \widetilde{H}_c(\pi) - \big[  \widetilde{\textbf{G}}_c(1) - \widetilde{\textbf{G}}_c(\pi) \big]^{-1} \big[ \widetilde{H}_c(1) - \widetilde{H}_c(\pi)  \big] \right\}
\end{align} 
Denote with $X = [ x_{t} I_{1t} \ \ x_{t} I_{2t} ] \equiv [ X_1 \ X_2 ]$ then the convergence of the covariance matrix 
\begin{align*}
\widetilde{M}_c( \pi )  := \left[ \mathcal{R} \left( X^{\prime} X \right)^{-1} \mathcal{R}^{\prime}\right] 
&=  \left[ \left( \frac{ X_1^{\prime} X_1}{ T^2 } \right)^{-1} + \left( \frac{ X_2^{\prime} X_2}{T^2} \right)^{-1} \right] \\
&\Rightarrow \left\{  \left( \int_0^{\pi} \ \widetilde{K}_c (r)  \widetilde{K}_c^{\prime} (r) dr \right)^{-1} + \left( \int_{\pi}^1 \widetilde{K}_c (r) \widetilde{K}_c^{\prime} (r) dr \right)^{-1}  \right\} \\
&\equiv 
\left\{ \widetilde{\textbf{G}}_c(\pi)^{-1} +  \left[ \widetilde{\textbf{G}}_c(1) - \widetilde{\textbf{G}}_c(\pi) \right]^{-1}  \right\}
\end{align*} 
Recall that the expression for the Wald statistic is as below
\begin{align}
\mathcal{W}_T^{OLS}( \pi ) 
&= 
\frac{1}{ \widehat{\sigma}_u^2 } \left( \widehat{\theta}_{1} -\widehat{\theta}_{2} \right)^{\prime}  \left[ \mathcal{R} \left( X^{\prime} X \right)^{-1} \mathcal{R}^{\prime}\right]^{-1} \left( \widehat{\theta}_{1} - \widehat{\theta}_{2} \right)
\end{align}
Therefore, we can now derive the limiting distribution of the sup OLS-Wald statistic in the case of the multiple predictive regression with persistent predictors. 
\begin{align*}
\widetilde{\mathcal{W}}^{OLS}( \pi ) 
&\Rightarrow  \underset{ \pi \in [ \pi_1 , \pi_2 ] }{ \text{ sup } }
\left\{  \widetilde{\textbf{G}}_c(\pi)^{-1}  H_c(\pi) - \big[ \widetilde{\textbf{G}}_c(1) - \widetilde{\textbf{G}}_c(\pi) \big]^{-1} \big[ \widetilde{H}_c(1) - \widetilde{H}_c(\pi)  \big] \right\}^{\prime} \\
& \ \ \ \ \ \ \ \ \ \ \ \  \ \ \times  \left\{\widetilde{\textbf{G}}_c(\pi)^{-1} +  \bigg[ \widetilde{\textbf{G}}_c(1) - \widetilde{\textbf{G}}_c(\pi) \bigg]^{-1}  \right\}^{-1} \\
& \ \ \ \ \ \ \ \ \ \ \ \ \ \ \times \left\{ \widetilde{\textbf{G}}_c(\pi)^{-1}   \widetilde{H}_c(\pi) - \big[ \widetilde{\textbf{G}}_c(1) -\widetilde{\textbf{G}}_c(\pi) \big]^{-1} \big[ \widetilde{H}_c(1) - \widetilde{H}_c(\pi)  \big] \right\}
\end{align*}
By applying the related inverse matrix formula to $\widetilde{\textbf{M}}_c( \pi )^{-1}$ we obtain that 
\begin{align*}
\widetilde{\textbf{S}}_c( \pi )^{-1} 
&\equiv 
\left\{ \widetilde{\textbf{G}}_c(\pi)^{-1} +  \bigg[ \widetilde{\textbf{G}}_c(1) -  \widetilde{\textbf{G}}_c(\pi) \bigg]^{-1}  \right\}^{-1} 
\\
&= 
\widetilde{\textbf{G}}_c(\pi) -\widetilde{\textbf{G}}_c(\pi) \bigg[ \widetilde{\textbf{G}}_c(\pi) +    \widetilde{\textbf{G}}_c(1) - \widetilde{\textbf{G}}_c(\pi) \bigg]^{-1} \widetilde{\textbf{G}}_c(\pi)
\\
&=  
\widetilde{\textbf{G}}_c(\pi) - \widetilde{\textbf{G}}_c(\pi) \widetilde{\textbf{\textbf{G}}}_c(1)^{-1}  \widetilde{\textbf{G}}_c(\pi)
\end{align*}
Thus, we show that the limiting distribution of the sup OLS-Wald statistic is given by
\begin{align}
\widetilde{\mathcal{W}}^{OLS}( \pi ) \equiv \underset{ \pi \in [ \pi_1 , \pi_2 ] }{ \text{ sup } } \ \mathcal{W}_T^{OLS}( \pi )  \Rightarrow \displaystyle \underset{ \pi \in [ \pi_1 , \pi_2 ] }{ \text{ sup } } \ \  \bigg\{ \widetilde{ \mathbf{N} }^{\prime}_c( \pi ) \widetilde{ \mathbf{M} }_c( \pi )^{-1}   \widetilde{ \mathbf{N} }_c( \pi ) \bigg\}
\end{align}
with quantities $\widetilde{ \mathbf{M} }_c( \pi )$, $\widetilde{ \mathbf{N} }_c( \pi )$,$  \widetilde{\textbf{G}}_c(\pi)$ and $\widetilde{H}_c(\pi)$ as defined by Proposition \ref{Proposition1}.
\end{proof}

\newpage

\textbf{Proof of Theorem \ref{Theorem2}.}  

\begin{proof}
Let $Y$ denote the vector with all demeaned values of $y_t$ and $X$ be the matrix collecting all demeaned values of $x_{t-1}$, that is, 
\begin{align*}
Y = \left( y_1 - \bar{y}_T, y_2 - \bar{y}_T, ... , y_T - \bar{y}_T  \right)^{\prime} \text{and} \ X = \left( x_0 - \bar{x}_{T-1}, x_1 - \bar{x}_{T-1}, ... , x_{T-1} - \bar{x}_{T-1} \right)^{\prime}.
\end{align*}
Similarly, we use $\mathcal{U}$ to denote the corresponding demeaned $u_{t}$ vector. Furthermore, for any $1 \leq t \leq T$, we define $X_t$ to be a $T \times p$ matrix, whose first $t$ rows are the same as $X$ while the rest are all zeros. Moreover, let $Z = \left( z_0, z_1,... z_{T-1} \right)^{\prime}$ collect all the IVX instruments, and $Z_t = \left( z_0, ..., z_t, 0,....,0 \right)^{\prime}$ be the corresponding time$-t$ truncated matrix. Given these notations, we express the original predictive regression model as 
\begin{align}
\label{model}
Y = X \beta_2 + X_t \phi + \mathcal{U}
\end{align} 
where $\phi := \beta_2 - \beta_1$ measures the magnitude of structural break. Moreover, we denote with $\phi_t$ to  the corresponding estimator which captures the break size associated with the sample partition at time $t$. Therefore, given any particular $t$, testing for structural break in the parameter vector $\beta$ is equivalent to testing the null hypothesis $\phi_t = 0$. Define with $\mathbf{M}_{xz} = \mathbf{I}_p - X \left( Z^{\prime} X \right)^{-1} Z^{\prime}$, which is idempotent and orthogonal to both $X$ and $Z$ and allows to rewrite \eqref{model} in its canonical form\footnote{Notice that the reparametrization of the model to its canonical form allows to shift the coordinates which transforms the model to its more general form within the exponential family.}. Multiplying $\mathbf{M}_{xz}$ on both sides of \eqref{model}, we deduce that $\mathbf{M}_{xz} Y = \mathbf{M}_{xz} X_t \phi_t + \mathbf{M}_{xz} \mathcal{U}$. Now, using $\mathbf{M}_{xz} Z_t$ as the instrumental variables for $\mathbf{M}_{xz}X_t$, we obtain an estimator for the parameter $\phi_t$ given by 
\begin{align}
\widetilde{\phi}_t = \left( Z_t^{\prime} \mathbf{M}_{xz} X_t \right)^{-1} Z_t^{\prime} \mathbf{M}_{xz} Y
\end{align}
Moreover, it holds that $\widetilde{\phi}_t = \widetilde{\beta}_2^{IVX} - \widetilde{\beta}_1^{IVX}$. Thus, the limiting distribution is given by
\begin{align*}
\widetilde{\phi}_t - \phi_t 
&=
\left( Z_t^{\prime} \mathbf{M}_{xz} X_t \right)^{-1} Z_t^{\prime} \mathbf{M}_{xz} \mathcal{U}
\\
&=
\bigg( Z_t^{\prime} X_t - Z_t^{\prime} X \left( Z^{\prime} X \right)^{-1} Z^{\prime} X_t  \bigg)^{-1} \bigg( Z_t^{\prime} \mathcal{U}_y - Z_t^{\prime} X_t \left( Z^{\prime} X \right)^{-1} Z^{\prime} \mathcal{U} \bigg)
\\
&=
\left[ \sum_{t=1}^{\floor{ T \pi } } \widetilde{z}_t \left( x_t - \bar{x}_{T-1} \right)^{\prime} - \sum_{t=1}^{\floor{T \pi} } \widetilde{z}_t \left( x_t - \bar{x}_{T-1} \right)^{\prime}  \left( \sum_{t=1}^{T} \widetilde{z}_t \left( x_t - \bar{x}_{T-1}  \right)^{\prime} \right)^{-1} \sum_{t=1}^{\floor{T \pi} } \widetilde{z}_t \left( x_t - \bar{x}_{T-1} \right)^{\prime} \right]^{-1}
\\
\nonumber
\\
& \times 
\left[ \sum_{t=1}^{\floor{T \pi} } \widetilde{z}_t \left( u_{t} - \bar{u}_{T} \right)^{\prime} - \sum_{t=1}^{\floor{T \pi} } \widetilde{z}_t \left( x_t - \bar{x}_{T-1} \right)^{\prime}  \left( \sum_{t=1}^{T} \widetilde{z}_t \left( x_t - \bar{x}_{T-1}  \right)^{\prime} \right)^{-1} \sum_{t=1}^{ T } \widetilde{z}_t \left( u_{t} - \bar{u}_{T} \right) \right] 
\end{align*}

\newpage 

Applying the asymptotic results given by Lemma \ref{lemma1} we obtain
\begin{align*}
T^{ \frac{ 1 + \gamma_x \wedge \delta_z }{2}} \left( \widetilde{\phi}_t - \phi_t \right) \Rightarrow \bigg[ \mathbf{\Psi} ( \pi ) - \mathbf{\Psi} ( \pi ) \mathbf{\Psi} ( 1 )^{-1} \mathbf{\Psi} ( \pi )^{\prime} \bigg]^{-1} \bigg( U(\pi) - \mathbf{\Psi} ( \pi ) \mathbf{\Psi} ( 1 )^{-1} U(1) \bigg)  
\end{align*} 
Next, we focus on covariance estimators for $\widetilde{\mathbf{Q} }_1(t) = \left( Z_t^{\prime} X_t \right)^{-1} \left( Z_t^{\prime} Z_t  \right) \left( X_t^{\prime} Z_t \right)^{-1}$ and the corresponding one for $\widetilde{\mathbf{Q}}_2(t)$. By ignoring the second-order degree bias correction we obtain:  
\begin{align}
\widetilde{\mathbf{Q}}_1(t) 
= 
\left( \sum_{t=1}^{\floor{ T \pi } } \widetilde{z}_t \left( x_t - \bar{x}_{T-1} \right)^{\prime} \right)^{-1} \left( \sum_{t=1}^{\floor{T \pi} } \widetilde{z}_t \widetilde{z}_t^{\prime}  \right) \left( \sum_{t=1}^{\floor{T \pi} } \left( x_t - \bar{x}_{T-1} \right) \widetilde{z}_t^{\prime} \right)^{-1}
\end{align}
which implies that
\begin{align}
T^{ 1 + \gamma_x \wedge \delta_z} \widetilde{\mathbf{Q}}_1(t) \Rightarrow \mathbf{\Psi}( \pi)^{-1} \left( \pi \sigma^2_u \widetilde{ \mathbf{V} } \right) \mathbf{\Psi}(  \pi)^{-1^{\prime}}
\end{align}
Similarly, 
\begin{align}
\widetilde{\mathbf{Q}}_2(t) 
= 
\left( \sum_{t=\floor{ T \pi } + 1}^{T} \widetilde{z}_t \left( x_t - \bar{x}_{T-1} \right)^{\prime} \right)^{-1} \left( \sum_{t=\floor{ T \pi } + 1}^{T} \widetilde{z}_t \widetilde{z}_t^{\prime}  \right) \left( \sum_{t=\floor{ T \pi } + 1}^{T} \left( x_t - \bar{x}_{T-1} \right) \widetilde{z}_t^{\prime} \right)^{-1}
\end{align}
which implies that
\begin{align}
T^{ 1 + \gamma_x \wedge \delta_z} \widetilde{\mathbf{Q}}_2(t) \Rightarrow \big( \mathbf{\Psi}( 1 ) - \mathbf{\Psi}(  \pi ) \big)^{-1} \left( (1 -  \pi ) \sigma^2_u \widetilde{\mathbf{V}} \right) \big( \mathbf{\Psi}( 1 ) - \mathbf{\Psi}(  \pi ) \big)^{-1^{\prime}}
\end{align}

Combining all the above we obtain the following result for the $\mathcal{W}_{b}(t)$ test statistic
\begin{align}
\mathcal{W}_{b}(t) 
&= \bigg( \widetilde{\beta}_2^{IVX}(t) - \widetilde{\beta}_1^{IVX}(t) \bigg)^{\prime} \left[ \widetilde{\mathbf{Q}}_1(t) + \widetilde{\mathbf{Q}}_2(t) \right]^{-1} \bigg( \widetilde{\beta}_2^{IVX}(t) - \widetilde{\beta}_1^{IVX}(t) \bigg)
\nonumber
\\
&= \left\{ T^{ \frac{ 1 + \gamma_x \wedge \delta_z}{2}} \bigg( \widetilde{\beta}_2^{IVX}(t) - \widetilde{\beta}_1^{IVX}(t) \bigg) \right\}^{\prime} \times \left[ T^{ 1 + \gamma_x \wedge \delta_z} \left( \widetilde{\mathbf{Q}}_1(t) + \widetilde{\mathbf{Q}}_2(t) \right) \right]^{-1} 
\nonumber
\\
&\times \left\{ T^{ \frac{ 1 + \gamma_x \wedge \delta_z}{2}} \bigg( \widetilde{\beta}_2^{IVX}(t) - \widetilde{\beta}_1^{IVX}(t) \bigg) \right\}
\end{align}
which implies the following weakly convergence result
\begin{align}
\label{expression.Wald1}
\mathcal{W}^{IVX}_{\beta}(t) 
&\Rightarrow \bigg( U( \pi ) - \mathbf{\Psi} ( \pi ) \mathbf{\Psi} ( 1 )^{-1} U(1) \bigg)^{\prime} \bigg[ \mathbf{\Psi} ( \pi ) - \mathbf{\Psi} ( \pi ) \mathbf{\Psi} ( 1 )^{-1} \mathbf{\Psi} ( \pi )^{\prime} \bigg]^{-1^{\prime}}
\nonumber
\\
&\times
\bigg[ \mathbf{\Psi}( \pi )^{-1} \left( \pi \sigma^2_y \widetilde{\mathbf{V}} \right) \mathbf{\Psi}( \pi )^{-1^{\prime}} + \big( \mathbf{\Psi}( 1 ) - \mathbf{\Psi}( \pi ) \big)^{-1} \left( (1 - \pi ) \sigma^2_y \widetilde{\mathbf{V}} \right) \big( \mathbf{\Psi}( 1 ) - \mathbf{\Psi}( \pi ) \big)^{-1^{\prime}}  \bigg]^{-1} 
\nonumber
\\
&\times
\bigg[ \mathbf{\Psi} ( \pi ) - \mathbf{\Psi} ( \pi ) \mathbf{\Psi} ( 1 )^{-1} \mathbf{\Psi} ( \pi )^{\prime} \bigg]^{-1} \bigg( U(\pi) - \mathbf{\Psi} (\pi ) \mathbf{\Psi} ( 1 )^{-1} U(1) \bigg) 
\end{align}

\newpage 

To simplify the notation of expression \eqref{expression.Wald1}, we denote with $A = \mathbf{\Psi}(\pi)$, $C = \mathbf{\Psi} (1)$ and $\Sigma = \sigma^2_u \widetilde{ \mathbf{V} }$. Then, we have the following equivalent form of the IVX-Wald statistic
\begin{equation}
\begin{aligned}
\mathcal{W}^{IVX}_{\beta}(t)  \equiv {} & \bigg( U(\pi) - A C^{-1} U(1) \bigg)^{\prime} \bigg( A - A C^{-1} A^{\prime} \bigg)^{-1^{\prime}} \\
      & \times \bigg[ \pi A^{-1} \Sigma A^{-1^{\prime}} + (1 - \pi) \big( C - A \big)^{-1} \Sigma \big( C - A \big)^{-1^{\prime}}  \bigg]^{-1} \\
      & \times \bigg( A - A C^{-1} A^{\prime} \bigg)^{-1} \bigg( U(\pi) - A C^{-1} U(1) \bigg)
\end{aligned}
\end{equation}
which can be written as below
\begin{equation}
\begin{aligned}
\mathcal{W}^{IVX}_{\beta}(t)  \equiv {} & \bigg( U(\pi) - A C^{-1} U(1) \bigg)^{\prime} \\
      & \times \bigg( \pi \left[ \bigg( A - A C^{-1} A^{\prime} \bigg) A^{-1} \Sigma A^{-1^{\prime}} \bigg( A - A C^{-1} A^{\prime} \bigg)^{\prime} \right]  \\
      & + (1 - \pi) \left[ \bigg( A - A C^{-1} A^{\prime} \bigg) \left( C - A\right)^{-1} \Sigma \left( C - A\right)^{-1^{\prime}} \bigg( A - A C^{-1} A^{\prime} \bigg)^{\prime} \right] \bigg) \\
      & \times \bigg( U(\pi) - A C^{-1} U(1) \bigg) \\
      & = \bigg( U(\pi) - A C^{-1} U(1) \bigg)^{\prime} \\  
      & \times \bigg( \pi \big( I - A C^{-1} \big) \Sigma \big( I - A C^{-1} \big)^{\prime} +  (1 - \pi) \left( A C^{-1} \right) \Sigma \left( A C^{-1} \right)^{\prime} \bigg)^{-1} \\
      & \times  \bigg( U(\pi) - A C^{-1} U(1) \bigg).   
\end{aligned}
\end{equation}

Notice that since $U(.)$ is known to be a Brownian motion with variance $\Sigma$, the above expression can be further simplified as following
\begin{equation}
\begin{aligned}
\label{equivalent.expression}
\mathcal{W}^{IVX}_{\beta}(t) \equiv {} & \bigg( B(\pi) - A C^{-1} B(1) \bigg)^{\prime} \\  
      & \times \bigg( \pi \big( I - A C^{-1} \big) \big( I - A C^{-1} \big)^{\prime} +  (1 - \pi) \left( A C^{-1} \right) \left( A C^{-1} \right)^{\prime} \bigg)^{-1} \\
      & \times  \bigg( B(\pi) - A C^{-1} B(1) \bigg).   
\end{aligned}
\end{equation} 

Substituting back to expression \eqref{equivalent.expression} the notation for  $A = \mathbf{\Psi}(\pi)$, $C = \mathbf{\Psi} (1)$ and $\Sigma = \sigma^2_u \widetilde{ \mathbf{V} }$, we obtain the expression below

\newpage 

\begin{equation}
\begin{aligned}
\mathcal{W}^{IVX}_{\beta}(t) \Rightarrow {} & \bigg( B(\pi) - \mathbf{\Psi} (\pi) \mathbf{\Psi} (1)^{-1} B(1) \bigg)^{\prime} \\  
      & \times \bigg( \pi \big( \mathbf{I}_p - \mathbf{\Psi} (\pi) \mathbf{\Psi} (1)^{-1} \big) \big( \mathbf{I}_p - \mathbf{\Psi} (\pi) \mathbf{\Psi} (1)^{-1} \big)^{\prime} +  (1 - \pi) \left( \mathbf{\Psi} (\pi) \mathbf{\Psi} (1)^{-1} \right) \left( \mathbf{\Psi} (\pi) \mathbf{\Psi} (1)^{-1} \right)^{\prime} \bigg)^{-1} \\
      & \times   \bigg( B(\pi) - \mathbf{\Psi} (\pi) \mathbf{\Psi} (1)^{-1} B(1) \bigg).   
\end{aligned}
\end{equation}  
where $B(.)$ is a standard Brownian motion. 

Using the asymptotic results given by Lemma \ref{lemma1}, we simplify expression $\mathbf{\Psi} (\pi) \mathbf{\Psi} (1)^{-1}$ to
\begin{equation}
\mathbf{R}( \pi ) := \mathbf{\Psi} (\pi) \mathbf{\Psi} (1)^{-1}
=
\begin{cases}
\displaystyle \left( \pi \mathbf{\Omega}_{xx} +  \int_0^{\pi} \underline{\mathbf{B}} dB^{\prime} \right) \left( \mathbf{\Omega}_{xx} +  \int_0^{1} \underline{\mathbf{B}} dB^{\prime} \right)^{-1}  
& ,\text{if} \ \gamma_x > 1
\\
\\
\displaystyle \left( \pi \mathbf{\Omega}_{xx} +  \int_0^{\pi} \underline{\mathbf{J}}_C dJ_C^{\prime} \right) \left( \mathbf{\Omega}_{xx} +  \int_0^{1} \underline{\mathbf{J}}_C dJ_C^{\prime} \right)^{-1}  
& ,\text{if} \ \gamma_x = 1
\\
\\
\displaystyle \pi \mathbf{I}_p
& ,\text{otherwise}. 
\end{cases}
\end{equation}
Therefore, by denoting the Brownian functional above with $\mathbf{N} (\pi) = \bigg( B(\pi) - \mathbf{R}(\pi) B(1) \bigg)$ and $\mathbf{M} (\pi ) = \bigg( \pi \left( \mathbf{I}_p - \mathbf{R}(\pi) \right)\left( \mathbf{I}_p - \mathbf{R}(\pi) \right)^{\prime} + (1 - \pi) \mathbf{R}(\pi) \mathbf{R}(\pi)^{\prime} \bigg)$, then we obtain 
\begin{align}
\widetilde{ \mathcal{W} }^{IVX}_{\beta}(t)  \Rightarrow \underset{ \pi \in [ \pi_1, \pi_2 ]}{\text{sup}} \bigg\{ \mathbf{N}( \pi )^{\prime} \mathbf{M}( \pi )^{-1} \mathbf{N}( \pi ) \bigg\}
\end{align}
\end{proof}

\textbf{Proof of Corollary \ref{corollary1}} When $\gamma_x < 1$, we have that $\mathbf{R} (\pi ) = \pi \mathbf{I}_p$ and thus $\mathbf{M} (\pi ) = \pi \left( \mathbf{I}_p - \pi \mathbf{I}_p \right)\left( \mathbf{I}_p - \pi \mathbf{I}_p \right)^{\prime} + (1 -\pi)\pi \mathbf{I}_p \left(\pi \mathbf{I}_p\right)^{\prime} = \left[ \pi (1 - \pi)^2 + (1 - \pi) \pi^2 \right] \mathbf{I}_p = \pi (1 - \pi) \mathbf{I}_p$. Hence, in this case, the limiting distribution in Theorem \ref{Theorem2} will reduce to 
\begin{align}
\widetilde{ \mathcal{W} }^{IVX}_{\beta}(t)  \Rightarrow  \underset{ \pi \in [ \pi_1, \pi_2 ]}{\text{sup}} \frac{ \big[ B(\pi) - \pi B(1) \big]^{\prime} \big[ B(\pi) - \pi B(1)\big] }{\pi (1 - \pi)}.
\end{align}

Thus, the above proof demonstrates that when we have predictors generated via the LUR specification with $\gamma_x < 1$ and testing the joint null hypothesis of no structural break and no predictability then the limiting distribution of the IVX Wald statistic converges to a NBB similar to the result of \cite{andrews1993tests} in the case of linear regression models.

\newpage 

\textbf{Proof of Corollary \ref{corollary2}.} 
\textit{(i)}. Using the above notations, it's straightforward to obtain that under the null hypothesis $\beta_1 = \beta_2 = 0$, we obtain that 
\begin{align}
\mathcal{W}_{ T}^{IVX}  = \widetilde{ \beta }^{IVX^{\prime}} \widetilde{\mathbf{Q}}_{\mathcal{R}}^{-1} \widetilde{\beta}^{IVX} \Rightarrow U(1)^{\prime} \left[ \mathbf{\Psi}(1)^{-1^{\prime}} \mathbf{\Sigma} \mathbf{\Psi}(1)^{-1} \right]^{-1} U(1).
\end{align}

Combining this result with Theorem \ref{Theorem2}, we can deduce that 
\begin{align*}
\mathcal{W}_{\beta}^{joint} &= \mathcal{W}_{ T}^{IVX} + \mathcal{W}_{\beta}^{IVX}(t) 
\nonumber
\\
&\Rightarrow 
\begin{pmatrix}
\mathbf{\Psi}(1)^{-1} U(1) 
\\
\\
U(\tau) - \mathbf{\Psi} ( \pi ) \mathbf{\Psi} ( 1 )^{-1} U(1) 
\end{pmatrix}^{\prime}
\begin{pmatrix}
\mathbf{\Psi}(1)^{-1^{\prime}} \mathbf{\Sigma} \mathbf{\Psi}(1)^{-1} \ &  \ \mathbf{0}_{p \times p}
\\
\\
\mathbf{0}_{p \times p} &  \mathbf{\Delta}_{p \times p}
\end{pmatrix}^{-1}
\begin{pmatrix}
\mathbf{\Psi}(1)^{-1} U(1) 
\\
\\
U(\pi) - \mathbf{\Psi} ( \tau ) \mathbf{\Psi} ( 1 )^{-1} U(1) 
\end{pmatrix}
\end{align*}
with
\begin{align*}
\mathbf{\Delta} := \pi \big( \mathbf{I}_p - \mathbf{\Psi} (\pi) \mathbf{\Psi} (1)^{-1} \big) \mathbf{\Sigma} \big( \mathbf{I}_p - \mathbf{\Psi} (\pi) \mathbf{\Psi}(1)^{-1} \big)^{\prime} +  (1 - \pi) \left( \mathbf{I}_p -  \mathbf{\Psi} (\pi) \mathbf{\Psi} (1)^{-1} \right) \mathbf{\Sigma} \left( \mathbf{I}_p -  \mathbf{\Psi} (\pi) \mathbf{\Psi} (1)^{-1} \right)^{\prime}
\end{align*}
Therefore, we obtain that 
\begin{align*}
\mathcal{W}_{\beta}^{joint} &= \mathcal{W}_{ T}^{IVX} + \mathcal{W}_{\beta}^{IVX}(t) 
\nonumber
\\
&\Rightarrow 
\begin{pmatrix}
\mathbf{\Psi}(1)^{-1} B(1) 
\\
\\
B(\pi) - \mathbf{\Psi} ( \pi ) \mathbf{\Psi} ( 1 )^{-1} B(1) 
\end{pmatrix}^{\prime}
\begin{pmatrix}
\mathbf{\Psi}(1)^{-1^{\prime}} \mathbf{\Psi}(1)^{-1} \ &  \ \mathbf{0}_{p \times p}
\\
\\
\mathbf{0}_{p \times p} &  \widetilde{\mathbf{\Delta}}_{p \times p}
\end{pmatrix}^{-1}
\begin{pmatrix}
\mathbf{\Psi}(1)^{-1} B(1) 
\\
\\
B(\pi) - \mathbf{\Psi} ( \pi ) \mathbf{\Psi} ( 1 )^{-1} B(1) 
\end{pmatrix}
\end{align*}
with
\begin{align*}
\widetilde{ \mathbf{\Delta} } := \pi \big( \mathbf{I}_p - \mathbf{\Psi} (\pi) \mathbf{\Psi} (1)^{-1} \big) \big( \mathbf{I}_p - \mathbf{\Psi} (\pi) \mathbf{\Psi} (1)^{-1} \big)^{\prime} +  (1 - \pi) \left( \mathbf{I}_p -  \mathbf{\Psi} (\pi) \mathbf{\Psi} (1)^{-1} \right) \left( \mathbf{I}_p -  \mathbf{\Psi} (\pi) \mathbf{\Psi} (1)^{-1} \right)^{\prime}
\end{align*}
which shows that 
\begin{align*}
\mathcal{W}_{\beta}^{joint} &= \mathcal{W}_{ T}^{IVX} + \mathcal{W}_{\beta}^{IVX}(t)  
\nonumber
\\
&\Rightarrow
\begin{pmatrix}
B(1) 
\\
\\
B(\pi) - \mathbf{R}( \pi ) B(1) 
\end{pmatrix}^{\prime}
\begin{pmatrix}
\mathbf{I}_p \ &  \ \mathbf{0}_{p \times p}
\\
\\
\mathbf{0}_{p \times p} &  \mathbf{M} ( \pi )
\end{pmatrix}^{-1}
\begin{pmatrix}
B(1) 
\\
\\
B(\pi) - \mathbf{R}( \pi ) B(1) 
\end{pmatrix}
\nonumber
\\
\nonumber 
\\
&\equiv B(1) B(1)^{\prime} + \mathbf{N}(\pi)^{\prime}\mathbf{M}( \pi )^{-1} \mathbf{N} ( \pi )
\end{align*}
Since the first component of the above decomposition is independent of $\pi$, then by the Continuous Mapping Theorem, we conclude that 
\begin{align*}
\widetilde{ \mathcal{W} }_{\beta}^{joint} = \underset{ \pi \in [ \pi_1, \pi_2 ] }{ \text{sup}  } \bigg\{ \mathcal{W}_{ T}^{IVX} + \mathcal{W}_{\beta}^{IVX}(t)   \bigg\} \Rightarrow B(1) B(1)^{\prime} + \underset{ \pi \in [ \pi_1, \pi_2 ] }{ \text{sup}  } \bigg\{ \mathbf{N}(\pi)^{\prime} \mathbf{M}( \pi )^{-1} \mathbf{N} (\pi ) \bigg\}.
\end{align*}

\newpage 

\textit{(ii)}. In case that $\gamma_x < 1$, by Corollary \ref{corollary1} it holds that the second component of the $\mathcal{W}_{\beta}$ test statistic reduces to a functional of a standard Brownian bridge $B(\pi) - \pi B(1)$. Then, since both $B(1)$ and $B(\pi) - \pi B(1)$ are Gaussian processes which implies that Cov$\big( B(1), B(\pi) - \pi B(1) \big) = \pi - \pi = 0$. Therefore, these two stochastic quantities are independent of each other\footnote{To validate the asymptotic independence of the two BM functionals, we can apply properties of the BM and prove that the covariance of the two terms is zero, which ensures independence.}. Hence, the proof of the statement follows. 

\textbf{Proof of Proposition \ref{Proposition1}.} Under the null hypothesis, $\mathbb{H}_0: \alpha_1 = \alpha_2$ and $\beta_1 = \beta_2 = \beta$, there is no break in the model intercept. In this section we consider in more details the estimator of the model intercept based on the IVX instrumentation before proving the asymptotic distribution given by Proposition \ref{Proposition1}. In particular, we propose to estimate the model intercept using the generated instrument instead of the predictor, we refer to this estimate as $\alpha^{IVZ}$ and the econometric intuition is explained below.

Based on the IVX estimation procedure, the corresponding IVX estimate for the model intercept is given by $\hat{\alpha} = \bar{y}_T - \widetilde{\beta}^{IVX} \bar{x}_{T-1}$. However, we notice that due to the presence of the predictors, then the limit theory for the estimate of the model intercept will vary with the degree of persistence of predictors.  We can see this below
\begin{align}
\sqrt{T} \left( \hat{\alpha}^{IVX} - \alpha \right) = \sqrt{T} \bar{u}_T -   \left[ T^{ \frac{1 + \gamma_x \wedge \delta_z }{2}} \left( \widetilde{\beta}^{IVX} - \beta \right) \right] \left[ T^{ - \frac{ \gamma_x \wedge \delta_z }{2}} \bar{x}_{T-1} \right], 
\end{align} 
where $\bar{u}_T = \frac{1}{T} \sum_{t=1}^T u_t$. Notice that both $\sqrt{T} \bar{u}_T$ and $T^{ \frac{1 + \gamma_x \wedge \delta_z }{2}} \left( \widetilde{\beta} - \beta \right)$ are both $\mathcal{O}_p(1)$, while the order of convergence of the last term depends on the persistence level of predictors. We have the following convergence rates
\begin{align}
\sum_{t=1}^T x_{t-1} = 
\begin{cases}
\mathcal{O}_p \left( T^{-1 / 2} \right)  
& ,\text{if} \ \gamma_x = 1
\\
\\
\mathcal{O}_p \left( T^{1 / 2 + \gamma_x} \right)    
& ,\text{if} \ 0 < \gamma_x < 1
\\
\\
\mathcal{O}_p \left( T^{ 3 / 2 } \right)  
& ,\gamma_x > 1. 
\end{cases}
\end{align}
We can observe that the term $T^{ - \frac{ \gamma_x \wedge \delta_z }{2}} \bar{x}_{T-1}$ will dominate in the limit in the case that $\gamma_x > \left( \delta_z + 1 \right) / 2$ and vanish if the reverse holds. In the case that $\gamma_x = \left( \delta_z + 1 \right) / 2$, all three terms appear in the asymptotic distribution which will depend on an unknown parameter $\beta$. To simplify the asymptotic theory we need to derive we estimate the model intercept based on the generated endogenous instrument, that is,  $\alpha^{IVZ} = \bar{y}_T - \widetilde{\beta}^{IVX} \bar{z}_{T-1}$, where $\bar{z}_{T-1} = \frac{1}{T-1} \sum_{j=2}^T \tilde{z}_{j-1}$. The advantage of the IVZ model estimate is that the persistence level of the instrument $\widetilde{z}_t$ is controlled by the choice of the tuning parameters $\delta_z$ and $c_z$ therefore the abstract degree of persistence is filtered out.  

\newpage 

Moreover, the  particular choice of the estimate for the model intercept works as a power enhancement mechanism against nonzero $\beta$ values, as seen below
\begin{align}
\label{ivz.intercept}
\left( \hat{\alpha}^{IVZ} - \alpha \right) =  \bar{u}_T - \left( \widetilde{\beta}^{IVX} - \beta \right) \bar{z}_{T-1} + \beta \left(  \bar{x}_{T-1} - \bar{z}_{T-1} \right), 
\end{align} 
The order of the first term is $\mathcal{O}_p \left( T^{-1 / 2}  \right)$.  Moreover, we can show that the second term is asymptotically dominated by the first term, while the third term does not converge unless the parameter space for $\beta$ is within the neighbourhood of zero. Therefore, power against the presence of predictability when a model intercept is included in the model is achieved when we control the convergence rate of the third term above.   

Next, we show that the second term of \eqref{ivz.intercept} is asymptotically dominated by the first term by expanding further the expression
\begin{align}
\label{second.term}
\sqrt{T} \left( \widetilde{\beta}^{\text{IVX}} - \beta \right) \bar{z}_{T-1} \equiv \left[ T^{ \frac{1 + \gamma_x \wedge \delta_z }{2}} \left( \widetilde{\beta}^{\text{IVX}} - \beta \right) \right] . \left[ T^{ - \left( 1 + \frac{\gamma_x \wedge \delta_z }{2} \right)}      \sum_{t=1}^T \widetilde{z}_{t-1} \right].
\end{align} 
The first term above is $\mathcal{O}_p (1)$ due to the convergence property of the IVX estimator to the mixed Gaussian distribution as established by \cite{phillipsmagdal2009econometric}. To establish the order of the second term we consider the convergence rate of the term $\sum_{t=1}^T \widetilde{z}_{t-1}$. For instance, by Lemma A2 in the Online Appendix of KMS we have that 
\begin{align}
\sum_{t=1}^T \widetilde{z}_{t-1} = 
\begin{cases}
\mathcal{O}_p \left( T^{ \frac{\gamma_x \wedge 1}{2} + \delta_z } \right)  
& ,\text{if} \ \delta_z < \gamma_z
\\
\\
\mathcal{O}_p \left( T^{ \gamma_x + \frac{\delta_z}{2} } \right)  
& , \text{if} \ 0 < \gamma_x < \delta_z. 
\end{cases}
\end{align}
Hence, we have that the second term of \eqref{ivz.intercept} will be $ \mathcal{O}_p \left( T^{ \frac{ \gamma_x \wedge 1 + \delta_z }{2} + 1} \right)$. Since $\gamma_x \wedge 1 + \delta_z < 2$, this implies that the order of the second term is $o_p (1)$. Therefore, we prove that the second term of $\left( \hat{\alpha}^{IVZ} - \alpha \right)$ is asymptotically dominated by the first term.  

For the third term of \eqref{ivz.intercept} we aim to show that is asymptotically dominated by the first term when $\beta$ is small enough. Thus, we rewrite with $\sqrt{T} \beta \left( \bar{x}_{T-1} - \bar{z}_{T-1} \right) = \beta \frac{1}{\sqrt{T}} \sum_{t=1}^T \left(   x_{t-1} - \widetilde{z}_{t-1} \right)$.  Using the representation formula of the instrument given by expression (23) of \cite{phillipsmagdal2009econometric}, we find an equivalent expression, that is, $\beta \frac{1}{\sqrt{T}} \sum_{t=1}^T \left(   x_{t-1} - \widetilde{z}_{t-1} \right) = - \beta C_z T^{ \left( \frac{1}{2} + \delta_z \right)} \sum_{t=1}^T \psi_{t-1}$, where $\psi_{t-1} = \sum_{j=1}^t R_T^{t-j} x_{j-1}$. Hence, if we set for simplicity $\mathbf{C}_z$ = $\mathbf{I}_p$ indicating a common degree of persistence across the instruments, we then obtain the following probability bound for this expression 
\begin{align*}
\norm{ \sqrt{T} \beta \left( \bar{x}_{T-1} - \bar{z}_{T-1} \right) } &\leq 
\norm{\beta } T^{- \left( \frac{1}{2} + \delta_z \right)} \sum_{t=1}^T \norm{ \psi_{t-1}} 
\\
&\leq  \norm{\beta} T^{- \left( \frac{1}{2} + \delta_z \right)} T \underset{ 2 \leq t \leq T}{ \text{sup}} \norm{\psi_{t-1}}
\end{align*}

\newpage 

\begin{align*}
&= \norm{ \beta } T^{ \left( \frac{1}{2} - \delta_z \right)}  \underset{ 2 \leq t \leq T}{ \text{sup}} \norm{\psi_{t-1}}
\\
&\leq \norm{ \beta } T^{ \left( \frac{1}{2} - \delta_z \right)} \mathcal{O}_p \left( T^{ \frac{\gamma_x}{2} + \delta_z }\right)
\\
&= \norm{ \beta } \mathcal{O}_p \left( T^{ \frac{\gamma_x}{2} + \delta_z }\right),
\end{align*}
Notice that the above result is justified due to the uniform bound of $\norm{ \psi_{t-1}}$, which is shown to be $\mathcal{O}_p \left( T^{ \frac{\gamma_x}{2} + \delta_z }\right)$ by PM. Thus, if $\beta = o_p \left( T^{- \frac{1 + \gamma_x }{2} } \right)$, then 
$\norm{ \sqrt{T} \beta \left( \bar{x}_{T-1} - \bar{z}_{T-1} \right)}= o_p(1)$ and the third component of $\left( \tilde{\alpha}^{IVZ} - \alpha \right)$ will also be asymptotically dominated. Therefore, when $\beta$ is a non-zero constant, this term will dominate, and this is the reason we obtain non-zero local power under the alternative hypothesis of predictability. 

\textbf{Proof of Proposition \ref{Proposition2}.}

For the proof of Proposition \ref{Proposition2}, we have already proved most of the required results.The only step we need to additionally show is that the test statistics $\mathcal{W}_{a}(t)$ and $\mathcal{W}_{b}(t)$ are asymptotically independent of each other. Notice that since the test statistic $\mathcal{W}_{a}(t)$ is driven by $\sum_{t=1}^{\floor{Tr}} u_t$, while the test statistic $\mathcal{W}_{b}(t)$ is driven by $\sum_{t=1}^{\floor{Tr}} \widetilde{z}_{t-1} u_t$. These two partial sums (invariance principles) have a joint weakly convergence to two independent Brownian motions, as shown by Proposition A1 in \cite{phillipsmagdal2009econometric}. Hence, the asymptotic independence is guaranteed. Then, convergence of the test statistic $\mathcal{W}_{ \alpha \beta}(t)$ follows by an application of the continuous mapping theorem.

\newpage


\newpage 

\section{Critical Values and Implementation}
\label{AppendixB}


\subsection{Critical Values}

The correct use of critical values close to the true asymptotic distribution of the test statistic under examination is crucial in  Monte Carlo simulations. This allows to correctly identify the existence of size distortions. Notice that for deciding whether to accept or reject the null hypothesis, one can use already tabulated critical values, only in those cases that the asymptotic distribution of the test statistic under consideration has an approximate form to Brownian functional (such as NBB). Moreover, we apply bootstrap methodologies (such as the fixed regressor bootstrap) to ensure that we obtain robust performance of the empirical size and power of the proposed tests.

\subsection{Software Implementation}

The Monte Carlo simulation study of the paper was implemented via the Statistical Package R (Version 3.5.1). More specifically, to validate the code for testing for both predictability and structural break, using the supremum Wald type statistics we utilize both the Matlab code of KMS as well as the \texttt{IVX} library written by \cite{Vasilopoulos2019} which provide implementations of the IVX instrumentation procedure. Moreover, to reduce the execution time we utilize parallel programming techniques as well as related R packages for this purpose, such as the \texttt{Rcpp} library and the \texttt{Rccp Armadillo} linear algebra library. Then, the critical values and empirical size and power of the tests were obtained by simulating the asymptotic distributions based on 10,000 Monte Carlo replications using the Iridis4 High Performance Computing Facility of the University of Southampton. 

\newpage 

\subsection{Bootstrap procedure}

For robustness of the procedure we use a bootstrap algorithm in order to obtain corresponding critical values. The Algorithm for the bootstrapped critical values that we follow in this paper is proposed by \cite{xu2020testing}, even though we do not consider the long-horizon predictability component of the particular procedure. We also consider the fixed regressor bootstrap as described by \cite{georgiev2018testing} and also in \cite{hansen2000testing}.

\paragraph{Algorithm 1.} \textcolor{red}{(to modify)}
Two-sided $100 \alpha \%$ Bootstrap Implied Test for $\mathbb{H}_0: R \theta = r_0$ 

Consider the predictive regression with multiple predictors given by 
\begin{align}
\label{m1}
y_t &= \mu_y + x_{t-1}^{\prime} \beta + u_{yt}, 
\\
\label{m2}
x_t &= \mu_x + R x_{t-1} + u_{xt}, 
\end{align}
where $R$ is a $k \times k$ dimensional coefficient matrix and the error sequence $u_t = \left( u_{yt},  u_{xt}^{\prime} \right)^{\prime}$ is $( k \times 1)-$dimensional martingale difference sequence with variance-covariance matrix 
\begin{align}
\text{Var}( u_t ) = \Sigma 
= 
\begin{bmatrix}
\sigma^2_y  & \ \sigma_{xy}^{\prime}
\\
\sigma_{xy}  & \ \Sigma_{xx}^{\prime}
\end{bmatrix}
\end{align} 

\begin{itemize}
\item[Step 1.] Estimation Step
\begin{itemize}
\item[1.1.] Run OLS regressions of \eqref{m1} and \eqref{m2}  to obtain the OLS estimates of the vector of coefficients $\left\{ \hat{\mu}_y, \hat{\beta}, \hat{\mu}_x, \hat{R}  \right\}$ and the sequence of residuals $\hat{u}_t = \left( \hat{u}_{yt},  \hat{u}_{xt}^{\prime} \right)^{\prime}$.

\item[1.2.] Compute the Wald statistic $\mathcal{W}_{IM}$.
\end{itemize}

\item[Step 2.] Generate the bootstrap sample $\left\{ y_t^{*}, x_t^{*}, t = 1,...,n \right\}$.
\begin{itemize}
\item[2.1.] Generate $u_t^{*} = \hat{e}_t e_t$, where $e_t$ is a random number generated from the standard normal distribution. Write $u_t^{*} = \left( u_{yt}^{*},  u_{xt}^{*\prime} \right)^{\prime}$.

\item[2.2.] Generate $x_t^{*}$ such that $x_t^{*} = \hat{\mu}_x + \hat{R}_{bc} x_{t-1}^{*} + u_{xt}^{*}$, where $\hat{R}_{bc}$ is the bias-corrected estimator of $R$, defined as below 
\begin{align*}
\hat{R}_{bc} = \hat{R} + \hat{\Sigma}_{xx} \bigg[ \big( I_k - \widehat{R}^{\prime} \big)^{-1} + \widehat{R}^{\prime} \big( I_k - \widehat{R}^{\prime 2} \big)^{-1} + \sum_{j=1}^k \widehat{\lambda}_j \big( I_k - \widehat{\lambda}_j \widehat{R}^{\prime} \big)^{-1} \bigg] \left( \sum_{t=2}^n \tilde{x}_{t-1} \tilde{x}_{t-1}^{\prime} \right),
\end{align*}
where $\widehat{\lambda}_j'$s for $j = 1,...,k$ are $k$ eigenvalues of $\widehat{R}^{\prime}$. We set $x_1^{*} = x_1$.

\item[2.3.] Generate $y_t^{*}$ such that $y_t^{*} = \hat{\mu}_y +  x_{t-1}^{* \prime} \hat{\beta} + u_{yt}^{*}$. We set $y_1^{*} = y_1$.

\item[2.4.] Repeat Steps 2.1 to 2.3 $B$ times (e.g., $B = 1000$).  
\end{itemize}

\newpage

\item[Step 3.] For each bootstrap sample, calculate the bootstrap Wald statistic
\begin{align*}
\mathcal{W}^{*}_{IM} = \left( \mathcal{R} \hat{\theta}^{*}_h - \mathcal{R} \hat{\theta}_h \right)^{\prime} \left[ \mathcal{R} \widehat{q}^{*} \left( \widehat{\Sigma}^{*} \otimes \left(  \sum_{t=2}^n \widetilde{x}^{*}_{t-1} \widetilde{x}^{*\prime}_{t-1} \right)^{-1}  \right) \widehat{q}^{* \prime} \mathcal{R}^{\prime} \right]\left( \mathcal{R} \hat{\theta}^{*}_h - \mathcal{R} \hat{\theta}_h \right),
\end{align*}
where $\hat{\theta}^{*}_h$, $\widehat{q}^{*}$, $\widehat{\Sigma}^{*}$ are computed like $\hat{\theta}_h$, $\widehat{q}$, $\widehat{\Sigma}$ except that the bootstrap sample $\left\{ y_t^{*}, x_t^{*} \right\}$ is replaced by the original sample $\left\{ y_t , x_t \right\}$.

\item[Step 4.] Use the $100 (1 - \alpha) \%$ quantile $\tau_{\mathcal{W}_{IM}^{*} } (1 - \alpha)$ of $\mathcal{W}_{IM}^{*}$ over $B$ bootstrap replications as the critical value, that is, the hypothesis $H_0$ is rejected at the significance level $\alpha$ if $\mathcal{W}_{IM} > \tau_{\mathcal{W}_{IM}^{*} } (1 - \alpha)$. 
\end{itemize}

\subsection{Structural Break Tests}

For the empirical application of the paper, we consider a set of additional structural break tests as robustness checks. Moreover, the particular structural break tests can be employed to identify the location of the break-point.

\newpage 

\section{Tables and Figures}
\label{AppendixC}

\subsection{Empirical Results}

\begin{table}[htbp]
  \centering
  \caption{Predictability Tests for the Equity Premium of univariate predictive regressions}
    \begin{tabular}{lccccccc}
    \hline    
    \hline
    Predictor & $\widehat{\beta}^{\text{OLS}}$ & $t^{\text{HAC}}$  & $R^2$    & $\widehat{\beta}^{\text{IVX}}$ & $\mathcal{W}^{\text{IVX}}$ & $sup\mathcal{W}^{\text{OLS}}$ & $sup\mathcal{W}^{\text{IVX}}$ \\
    \hline
    Panel A: & \multicolumn{7}{l}{1946Q1 - 2019Q4} \\
    \hline
    Dividend payout ratio & 0.0056 & 0.3898 & 0.0003 &       &       &       &  \\
    Long-term yield & -0.0824 & -1.5958 & 0.0032 &       &       &       &  \\
    Dividend yield  & 0.0150 & 1.9880 & 0.0047 &       &       &       &  \\
    Dividend-price-ratio & 0.0142 & 1.8741 & 0.0042 &       &       &       &  \\
    T-bill rate & -0.1043 & -2.1461$^{*}$ & 0.0060 &       &       &       &  \\
    Earnings-price-ratio & 0.0115 & 1.2885 & 0.0028 &       &       &       &  \\
    Book-to-market ratio & 0.0042 & 0.6590 & 0.0006 &       &       &       &  \\
    Default yield spread & 0.0150 & 1.9880$^{*}$ & 0.0047 &       &       &       &  \\
    Net equity expansion & -0.0501 & -0.6149 & 0.0005 &       &       &       &  \\
    Term spread & 0.1813 & 1.6834 & 0.0035 &       &       &       &  \\
    Inflation rate & -0.9075 & -2.6520$^{**}$ & 0.0096 &       &       &       &  \\
    \hline
    Panel B: & \multicolumn{7}{l}{1990Q1 - 2019Q4} \\
    \hline
    Dividend payout ratio & 0.0028 & 0.1597 & 0.0001 &       &       &       &  \\
    Long-term yield & -0.0883 & -0.8409 & 0.0017 &       &       &       &  \\
    Dividend yield  & 0.0354 & 1.6497 & 0.0102 &       &       &       &  \\
    Dividend-price-ratio & 0.0351 & 1.6389 & 0.0101 &       &       &       &  \\
    T-bill rate & -0.0393 & -0.3885 & 0.0005 &       &       &       &  \\
    Earnings-price-ratio & 0.0173 & 0.8403 & 0.0041 &       &       &       &  \\
    Book-to-market ratio & 0.0317 & 1.0439 & 0.0041 &       &       &       &  \\
    Default yield spread & 0.5275 & 0.5506 & 0.0025 &       &       &       &  \\
    Net equity expansion & 0.1188 & 0.9655 & 0.0037 &       &       &       &  \\
    Term spread & -0.0727 & -0.4442 & 0.0005 &       &       &       &  \\
    Inflation rate & 0.2005 & 0.2831 & 0.0003 &       &       &       &  \\
    \hline
    \hline
    \end{tabular}%
  \label{tableB1}%
\end{table}%

\begin{footnotesize}
Table \ref{tableB1} presents simple predictability tests of the null hypothesis $\beta = 0$ and model estimates based on the predictive regression with a single predictor given by $y_t = \alpha + \beta x_{t-1} + u_t$, applied to all individual predictors. For the t-tests we consider a t-ratio based on the HAC (Newey-West) covariance estimator. For all test statistics, we denote the rejection probabilities under the null hypothesis of no predictability at significance levels $1\% ( ^{***} )$, $5\% ( ^{**} )$, and $10\% ( ^{*} )$, respectively. 
\end{footnotesize}


\newpage

\begin{table}[htbp]
  \centering
  \caption{Structural Break Tests for predictors}
    \begin{tabular}{lcccccc}
    \hline
    \hline
    Panel A: & \multicolumn{6}{l}{1990Q1 - 2019Q4} \\
    \hline
    Predictor &       & $\widehat{\rho}$   & $R^2$    & BP Seq F-test & Max LR F-test & Exp LR F-test \\
    \hline
    Dividend payout ratio &       & 0.7477 & 0.5592 & 4.0061 & 49.4450$^{***}$ & 19.1970$^{***}$ \\
    Long-term yield &       & 0.0429 & 0.0046 & 3.4002 & 2.3062 & 0.3676 \\
    Dividend yield  &       & 0.0424 & 0.0001 & 3.5761 & 7.2497 & 1.4443 \\
    Dividend-price-ratio &       & 0.0511 & 0.0008 & 3.6305 & 7.5262 & 1.5400 \\
    T-bill rate &       & 0.4734 & 0.2169 & 2.6279 & 5.4363 & 0.6033 \\
    Earnings-price-ratio &       & 0.4385 & 0.1922 & 1.6033 & 14.6679$^{***}$ & 4.6243$^{***}$ \\
    Book-to-market ratio &       & -0.0439 & 0.0012 & 5.5639 & 3.1215 & 0.1470 \\
    Default yield spread &       & 0.4616 & 0.2131 & 3.1040 & 7.4651 & 1.9635$^{**}$ \\
    Net equity expansion &       & 0.2440 & 0.0596 & 0.9001 & 2.5452 & 0.2197 \\
    Term spread &       & 0.1035 & 0.0107 & 1.6129 & 2.1243 & 0.2136 \\
    Inflation rate &       & -0.0979 & 0.0096 & 3.2371 & 5.1032 & 0.9540 \\
    \hline
    Predictor & $\widehat{\mu}$ & $\widehat{\rho}$ & $R^2$    & BP Seq F-test & Max LR F-test & Exp LR F-test \\
    \hline
    Dividend payout ratio & -0.0001 & 0.7476 & 0.5592 & 26.5211$^{***}$ & 26.5211$^{***}$ & 7.7360$^{***}$ \\
    Long-term yield & -0.0002 & 0.0368 & 0.0014 & 1.7122 & 1.1633 & 0.1840 \\
    Dividend yield  & -0.0007 & 0.0408 & 0.0017 & 3.4744 & 5.0447 & 1.1532 \\
    Dividend-price-ratio & -0.0007 & 0.0495 & 0.0024 & 3.3367 & 5.0907 & 1.1876 \\
    T-bill rate & -0.0001 & 0.4684 & 0.2197 & 1.3134 & 2.6698 & 0.4553 \\
    Earnings-price-ratio & -0.0003 & 0.4384 & 0.1922 & 0.7989 & 7.5124$^{***}$ & 2.2289 \\
    Book-to-market ratio & -0.0006 & -0.0447 & 0.0020 & 2.7760 & 2.0446 & 0.2170 \\
    Default yield spread & 0.0000 & 0.4616 & 0.2131 & 1.6369 & 3.7191 & 0.7621 \\
    Net equity expansion & 0.0000 & 0.2440 & 0.0596 & 0.8209 & 1.6173 & 0.1941 \\
    Term spread & 0.0000 & 0.1034 & 0.0107 & 1.3025 & 1.7316 & 0.2593 \\
    Inflation rate & 0.0000 & -0.0979 & 0.0097 & 1.6209 & 2.5489 & 0.4311 \\
    \hline
    \hline
    \end{tabular}%
  \label{tableB2}%
\end{table}%

\begin{footnotesize}
Table \ref{tableB2} presents structural break tests and model estimates based on an AR(1) model applied to all individual stationary predictors (based on first differences, i.e., $\Delta x = x_t - x_{t-1}$) for Panel A with sampling period: 1990Q1 - 2019Q4. The covariance matrix for the AR(1) model is constructed using the HAC (Newey-West) estimator. We consider the following structural break tests\footnote{Further details regarding the specification and  testing algorithms can be found in the User's Guide of Eviews  under the section titled "Stability Diagnostics", see \url{ https://www.eviews.com/help/}} (i) Bai-Perron F-test based on  the sequential break detection algorithm of \cite{bai2003computation} with $\epsilon = 0.15$; (ii) Maximum LR F-statistic with $\epsilon = 0.15$; and (iii) Exp LR F-statistic with $\epsilon = 0.15$. The last two structural break tests represent the Andrews unknown breakpoint tests (see, \cite{andrews1993tests}). Notice also for the BP statistic we assume a common data distribution across the blocks to ensure consistent estimation of the variance. For all test statistics, we denote the rejection probabilities under the null hypothesis of no structural break at significance levels $1\% ( ^{***} )$, $5\% ( ^{**} )$, and $10\% ( ^{*} )$, respectively. 
\end{footnotesize}

\newpage

\end{document}